\documentclass[preprint]{elsarticle}
\usepackage[margin=2.5cm]{geometry}
\usepackage{amssymb}
\usepackage{lineno}
\usepackage{graphicx,caption}
\usepackage{amsmath} 
\usepackage{booktabs,multirow,array}
\usepackage{rotating}
\usepackage[table, dvipsnames]{xcolor}
\usepackage[font=small]{subcaption}
\usepackage{setspace}
\usepackage{etoolbox,siunitx}
\usepackage{epsfig}
\usepackage{url}
\usepackage{makecell}

\journal{ISPRS Journal of Photogrammetry and Remote Sensing }

\begin{document}
	
\captionsetup[figure]{font=small}

\begin{frontmatter}

\title{Fusion of TanDEM-X and Cartosat-1 Elevation Data Supported by Neural Network-Predicted Weight Maps}

\author[1]{Hossein Bagheri}
\author[1]{Michael Schmitt}
\author[1,2]{Xiao Xiang Zhu}

\address[1]{Signal Processing in Earth Observation, Technical University of Munich, Munich, Germany}
\address[2]{Remote Sensing Technology Institute, German Aerospace Center, Oberpfaffenhofen, Wessling, Germany
}

\begin{abstract}
\textcolor{Red}{This is the pre-acceptance version, to read the final version, please go to ISPRS Journal of Photogrammetry and Remote Sensing on ScienceDirect}. Recently, the bistatic SAR interferometry mission TanDEM-X provided a global terrain map with unprecedented accuracy. However, visual inspection and empirical assessment of TanDEM-X elevation data against high-resolution ground truth illustrates that the quality of the DEM decreases in urban areas because of SAR-inherent imaging properties.
One possible solution for an enhancement of the TanDEM-X DEM quality is to fuse it with other elevation data derived from high-resolution optical stereoscopic imagery, such as that provided by the Cartosat-1 mission. This is usually done by Weighted Averaging (WA) of previously aligned DEM cells. 
The main contribution of this paper is to develop a method to efficiently predict weight maps in order to achieve optimized fusion results. The prediction is modeled using a fully connected Artificial Neural Network (ANN). The idea of this ANN is to extract suitable features from DEMs that relate to height residuals in training areas and then to automatically learn the pattern of the relationship between height errors and features. The results show the DEM fusion based on the ANN-predicted weights improves the qualities of the study DEMs. Apart from increasing the absolute accuracy of Cartosat-1 DEM by DEM fusion, the relative accuracy (respective to reference LiDAR data) of DEMs is improved by up to 50\% in urban areas and 22\% in non-urban areas while the improvement by the HEM-based method does not exceed 20\% and 10\% in urban and non-urban areas respectively.     

\end{abstract}


\end{frontmatter}


\section{Introduction}\label{sec:Introduction}

Digital Elevation Models (DEMs) in diverse resolutions, levels of height accuracy and coverages are routinely produced by different techniques for a varied range of applications in different fields, such as navigation, geographical studies of the environment, or the ortho-rectification of remote sensing imagery. Particular attention is paid to the production of global DEMs, which represent homogeneous topography information for nearly all landmasses of the world. Different technologies have been employed for producing nearly global DEMs like the SRTM DEM \cite{Rabus2003},\cite{Rodriguez2006}, the ASTER GDEM \cite{Tachikawa2011} or AW3D30 \cite{takaku2014generation,tadono2014precise} which methodologically lie in two categories: SAR-interferometric and optical stereoscopic procedures. 
Each one of them has its own advantages and drawbacks that lead to DEMs with specific properties and limitations regarding final resolution and coverage.
As an example, the SRTM DEM with a grid spacing of 1$''$ only covers the latitudes between 56$^{\circ}$S and 60$^{\circ}$N. An example for an elevation model derived from optical stereo data is the AW3D30 DEM based on ALOS PRISM data, which provides both higher accuracy and larger coverage (between 82$^{\circ}$S and 83$^{\circ}$N) than the SRTM DEM, but contains some void areas due to missing information caused by clouds, snow etc. \cite{takaku2016validation}.  

Recently, a new global topography dataset was attained through the TanDEM-X mission, which provides a  spatial resolution of 12 m with coverage of nearly the whole earth. The TanDEM-X mission comprises twin SAR satellites (TerraSAR-X and TanDEM-X launched in June 2007 and June 2010, respectively), which fly in adjacent orbits to acquire bistatic SAR images. The mission was devised to produce DEMs with a target accuracy according to High-Resolution Terrain Information standard level 3 (HRTI-3) \cite{heady2009high}: i.e., with a relative height accuracy finer than 2 m for areas including slopes lower than 20\%,  and 4 m for slopes steeper than 20\% \cite{Krieger2007}. 
The special satellite constellation equipped with X-band SAR sensors exploits a bistatic SAR interferometry configuration with single pass acquisitions free of atmospheric and temporal decorrelation effects and consequently provides the first high-resolution global DEM. The initial DEM product, the so-called raw TanDEM-X DEM with nominal pixel spacing of 0.2 arcsec (6 m at the equator), is the output of the Integrated TanDEM-X Processor (ITP) \cite{6049701}.
The raw TanDEM-X DEM is finally cast in a grid with pixel spacing of 0.4 arcsec (12 m at the equator) after DEM calibration \cite{Gruber2012},\cite{Rossi2012} and mosaicking \cite{Gruber2016} to obtain the global DEM according to HRTI-3 standard. While the standard DEM globally represents non-urban areas with unprecedented relative accuracy \cite{Zink2014}, the drop of the DEM's spatial resolution makes the final standard DEM unsuitable for high-resolution 3D reconstruction in urban areas \cite{Rossi2013}. Consequently, the raw TanDEM-X DEM provides a more spatially detailed mapping of urban areas in comparison to the standard version of the global TanDEM-X DEM.  However, preliminary visual inspection of raw TanDEM-X DEM data still indicates unfavorable spatial resolution and drop of height precision, especially for areas with topographically difficult surfaces \textemdash like urban areas \cite{5764720}\textemdash and the requirement for TanDEM-X quality enhancement in these areas.

One solution for refining the TanDEM-X DEM in difficult terrains can be a fusion with elevation data derived from other sources with different acquisition properties.  Optical imagery, e.g., does not suffer from SAR-intrinsic imaging effects, such as layover and shadowing, which influence the appearance of InSAR DEM products.   

Optical DEMs result from stereoscopic 3D reconstruction of high-resolution optical images. For example, Cartosat-1 data provides a series of DEMs with relative accuracy of HRTI-3 standard (2-3 m). Cartosat-1 (also called IRS-P5) is an Indian satellite (launched in May 2005) equipped with a pushbroom sensor consisting of an ensemble of CCDs with a size of 2.5 m in two lines for along track scanning of scenes with a stereo angle of 31$^{\circ}$ \cite{Srivastava2007}. It is particularly intended to produce a high-resolution DEM  with coverage of a relatively wide area \cite{Ahmed2007}, and is used, for instance, for large-scale DEM generation in Europe \cite{Uttenthaler2013} . The Cartosat-1 data are provided with Rational Polynomial Coefficients (RPCs) computed from the mission's orbit and attitude information. Evaluations have demonstrated that their accuracy -for instance measured by Root Mean Square Errors (RMSE)- is restricted to multiple hundred meters \cite{Lehner2007} i.e. the final produced DEM in spite of fairly high relative accuracy is absolutely located in an incorrect position. The poor accuracy of the RPCs affects the stereo intersection results and causes residuals in the final DEM product.  
Generally, a good distribution of Ground Control Points (GCPs)  is needed for RPC refinement and bias compensation \cite{Teo2011} of high-resolution optical images like those provided by Cartosat-1, but availability of GCPs cannot always be ensured.
The conventional solution is to use available global DEMs\textemdash like the SRTM DEM \textemdash as an external vertical reference for bias compensation and RPC refinement \cite{Kim2011}. The emergence of the TanDEM-X DEM as global DEM of HRTI-3 standard, as opposed to the SRTM DEM of DTED-2 standard \cite{Krieger2007}, provides the required height reference with higher accuracy for refining the RPCs of Cartosat-1 that will ultimately result in more accurate Cartosat-1 DEM. 

Considering the aforementioned defects of the Cartosat-1 and TanDEM-X elevation data, the main objective of this paper is to develop a framework for efficient fusion of TanDEM-X and Cartosat-1 DEM over urban areas. Eventually, this fusion will increase the height precision of the final DEM over urban areas, while its absolute vertical accuracy is improved to the level of the TanDEM-X DEM.

Data fusion approaches with great deal of applications in remote sensing can be adapted for DEM fusion tasks \cite{7740215}, and for this aim, various methods have been investigated for different kinds of DEMs.
Reinartz et al. \cite{Reinartz2005} employed weighted averaging for the fusion of SPOT-5 and SRTM DEMs. In a similar study \cite{Roth2002}, weighted averaging was used to fuse ERS TanDEM data and SRTM data with MOMS-2P data. A more advanced technique was proposed by Papasaika \cite{Papasaika:2011:FDE:2050390.2050409}, in which sparse representation supported by weights served for fusion of DEMs from various data sources. Pock et al. \cite{Pock2011} proposed Total Generalized Variational (TGV) methods for fusion of airborne optical-stereoscopic DEMs, while a weighted version of total variational (TV) method and TGV were examined by Kuschk et al. \cite{7752839} on different space borne optical DEMs. Fuss et al. utilized the modified K-means clustering algorithm to fuse multiple overlapping radargrammetric Envisat-2 DEMs \cite{Fuss2016}. The first experience for fusion of TanDEM-X and Cartosat-1 DEMs over urban and non-urban areas comes with \cite{dlr87859}, in which only the TanDEM-X DEM was improved over non-urban areas by weighted averaging and prior knowledge of DEM qualities. 

Among the aforementioned methods, weighted averaging (WA) is wellknown and frequently used for DEM fusion purposes \cite{Gruber2016},\cite{Reinartz2005},\cite{Roth2002},\cite{dlr87859},\cite{Bagheri2017},\cite{Deo2015}, because of its simple implementation and low computational cost. In addition, most advanced techniques apply weights to assist the fusion process to reach the desired output. This means the weights play a key role for efficient fusion of DEMs, especially in the case of multi-sensor DEM fusion, like stereoscopic-optical and InSAR DEMs \cite{Bagheri2017}.
The key problem with using DEM fusion approaches, especially for WA, is applying appropriate weight maps receptive to each DEM\textemdash which used to be proportional to the expected height residuals. For this purpose, prior knowledge about existing DEM errors will always be beneficial for the fusion process. One solution for predicting the expected errors is based on an error propagation analysis through the DEM generation procedure. However, usually such a model can only be an approximation and may not model all potential error sources.

An alternative is to learn the error patterns by comparing exemplary areas of interest and corresponding ground truth reference data: e.g., derived from high-precision LiDAR measurements. This way, suitable weights can be predicted for newly incoming datasets for which neither detailed information about the height errors nor any ground truth data are available.   

This paper is an extension of \cite{Bagheri2017,bagheri2017fusion}, in which we mostly focused on evaluating the accuracy of Cartosat-1 and TanDEM-X DEMs with respect to high-resolution LiDAR reference data to provide a judgment about the potential of the Cartosat-1 and TanDEM-X DEM fusion over urban areas.
The evaluation confirmed that the TanDEM-X DEM quality over urban areas is not ideal in comparison to the Cartosat-1 elevation data, and that data fusion can improve the quality of the final DEM product.

In this paper, a sophisticated framework for appropriate weight map prediction is proposed (Section \ref{subsec:ANN-WAfusion}). Firstly, suitable spatial features, along with height residuals, are extracted from the training DEMs. After that, the pattern of refined height errors in relation to features are learned by an artificial neural network (ANN) to predict weights for WA DEM fusion. In order to provide a baseline result, first simple DEM fusion using the weights derived from the TanDEM-X Height Error Map (HEM) and the Cartosat-1 matching standard deviations are shown and discussed in Section \ref{sec:DEMfusion} and Section \ref{sec:fusionresult}. The final results (Section \ref{sec:fusionresult}) illustrate the ANN-supported DEM fusion can improve the quality of both DEMs over urban areas to generate a global high-resolution DEM in contrast to standard HEM-based weights. Finally, fusion framework is validated using SRTM DEM and the ASTER GDEM data to investigate the possibility of transferring the proposed approach to DEMs of other specifications as well (Section \ref{Extensionfusion}).

\section{Study Area and DEMs}\label{subsec:studyArea}
The data used for the experiments described in this paper were acquired over the area of Munich, Germany. The Cartosat-1 DEM with nominal grid size of 5 m was generated from stacks of overlapping images by the dense matching and 3D stereoscopic reconstruction toolboxes embedded in the XDibias image processing system of DLR. The description of the DEM generation procedure has been detailed in \cite{dlr55978}. The TanDEM-X raw DEM produced by DLR's Integrated TanDEM-X Processor (ITP), in which bistatic SAR data-takes acquired in strip map mode are processed interferometrically. It is delivered with a grid spacing of 0.2 arc seconds. More details about the used TanDEM-X and Cartosat-1 tiles are collected in Tabs \ref{PropertyTDX} and \ref{PropertyCS1}.

\begin{table*}[h]
	\centering
	
	\begin{tabular}{l c }
		\hline
		\multicolumn{2}{c}{Raw TanDEM-XDEM} \\\hline
		Center incidence angle      &38.25$^{\circ}$  \\
		Equator crossing direction  &Ascending   \\
		Look direction              & Right   \\
		Height of ambiguity   &45.81m           \\
		Total number of looks       &22         \\
		Pixel spacing               &0.2 arcsec  \\
		HEM mean                    &1.33 m     \\
		\hline	
			
	\end{tabular} 
	\caption{Properties of raw TanDEM-X tile.}\label{PropertyTDX}
\end{table*}

\begin{table*}[h]
	\centering
	
	\begin{tabular}{l c }
		\hline
		\multicolumn{2}{c}{Cartosat-1 DEM}\\\hline
		Stereoscopic angle   &31$^{\circ}$ \\
		Max number of rays   &11\\
		Min number of rays   &2\\
	    Horizontal reference&BKG orthophotos$^{*}$ \\
		Vertical reference   & SRTM DEM\\
		Pixel spacing       &5 m\\
		 Mean height error (1$\sigma$)  &2-3 m\\
		\hline	
		
	\end{tabular} 
	\caption{Properties of Cartosat-1 tile. $^{*}$ For more information, look up \cite{Cartography} }\label{PropertyCS1}
\end{table*}

\begin{figure}[ht!]
	\begin{center}
		\includegraphics[width=1\textwidth]{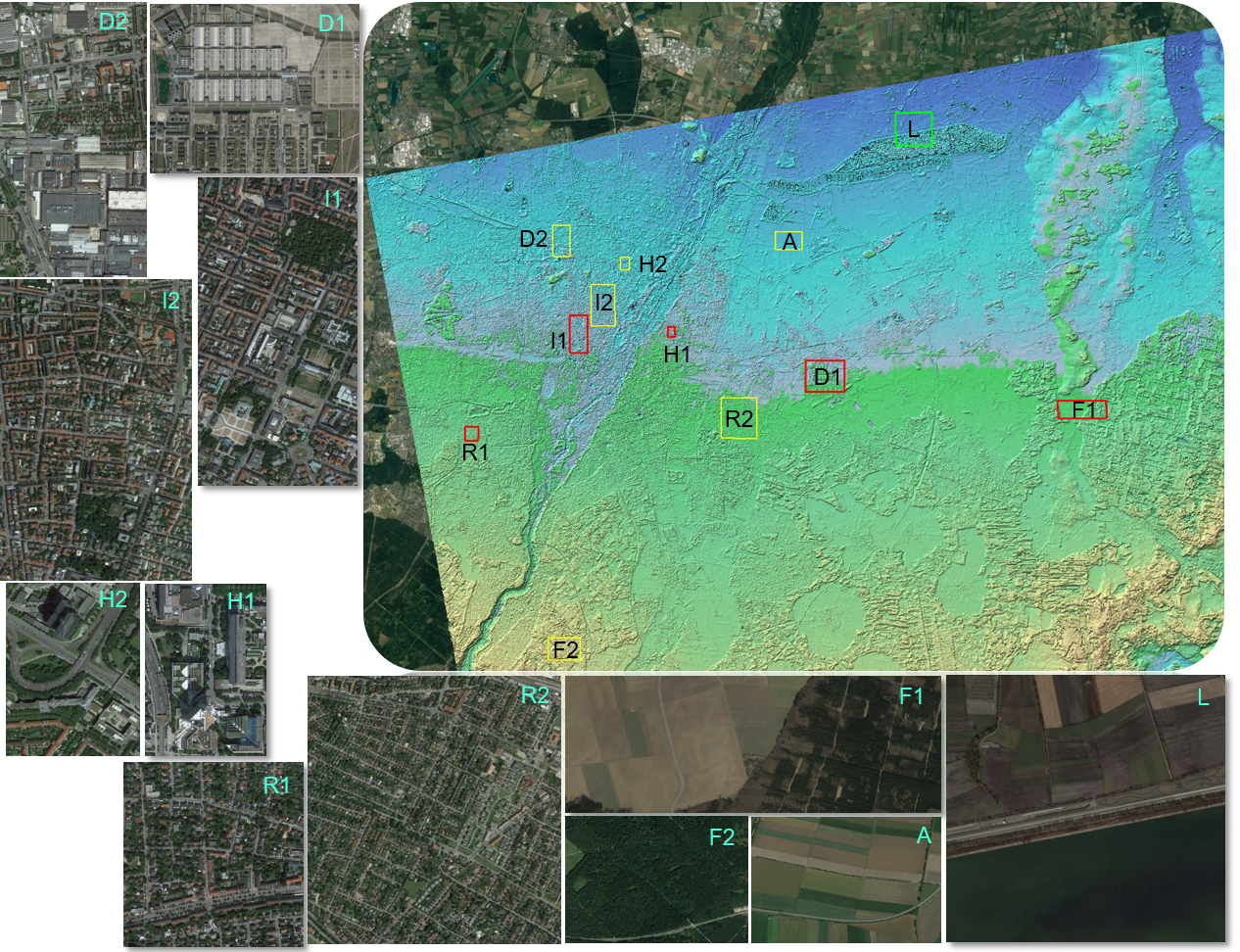}
		\caption{Display of land types and location of study subsets from Munich area}
		\label{studyarea}
	\end{center}
\end{figure}

For representative experiments, we chose study subsets representing 6 different land types that are usually found over urban areas and their surroundings (see Fig\ref{studyarea}). The main characteristics of these land types are briefly stated in Tab \ref{studyareades}.  The naming abbreviations are meant to facilitate referencing each subset throughout this paper. 

\begin{table*}[h]
	\centering
	
	\begin{tabular}{l c c c}
		\hline
		\multicolumn{3}{c}{Descriptions of study areas}\\
		Subsets   &Characteristics& Naming & Areas ($km^{2}$) \\\hline
		Industrial   &open mid rise rectangular buildings&D1, D2 & 1.77, 0.78 \\
		Inner city   &compact  mid rise with complicated shape  & I1, I2 & 0.92, 1.384\\
		&and perhaps relatively high&\\
		High building &high rise buildings and skyscrapers&H1, H2 & 0.09, 0.15\\
		Residential       & open low rise buildings for single family &R1, R2& 0.26, 2.12\\
	Forested    &dense canopy of trees& F1, F2 & 0.54, 1.0 \\
	Agricultural    &classical land farms &F1, A& 0.68, 0.66\\
		\hline	
		
	\end{tabular} 
	\caption{The main properties of study subsets}\label{studyareades}
\end{table*}

Both for  training and evaluation of the ANN-fusion framework, highly accurate reference elevation models are provided by high-resolution LiDAR point clouds with a density of one point per square meter. 

\section{Data Preparation and Alignment}\label{subsec:DEMalign}

Before implementing the fusion process, the Cartosat-1 and TanDEM-X data should be homogenized in terms of horizontal and vertical references, as well as pixel spacing. The initial coordinate systems and the pixel spacing of the study DEMs are expressed in Tab \ref{DEMspeci}. The final DEM specifications identify the reference systems and nominal pixel spacing of all used DEMs after the data preparation process. To prepare the data for the actual fusion, the Cartosat-1 and the TanDEM-X DEMs are resampled to 5 m pixel spacing, which is nearly identical to their own initial resolutions and grid spacing.

\begin{table*}[h]
	\centering
	\begin{tabular}{l c c c}
		\hline
		\multicolumn{1}{c}{}&\multicolumn{1}{c}{Horizontal reference}&\multicolumn{1}{c}{Vertical reference} &\multicolumn{1}{c}{Pixel spacing}\\\hline
		{Cartosat-1}    &UTM(WGS84)     &EGM96	&5m\\
		{TanDEM-X}      &WGS84          &WGS84 	&0.2 arcsec($\sim$6m)\\
		{LiDAR}         &Gauss Kr\"{u}ger(Bessel)  &Bessel	&1m\\
		{Final DEM}     &UTM(WGS84)     &WGS84	&5m\\
		\hline		
	\end{tabular} 
	\caption{Specifications of initial DEMs and final DEM, output of preparation step. }\label{DEMspeci}
\end{table*} 

In addition, after preparing the elevation data in the form of the desired grid and reference, the tiles of study DEMs (TanDEM-X and Cartosat-1) should be aligned together to compensate any rotational and translational discrepancies. Usually, the best operational way for this purpose would be to utilize the well calibrated TanDEM-X data as an external DEM for a refinement of the Cartosat-1 RPCs. However, since this study does not focus on Cartosat-1 DEM generation and only starts with an available Cartosat-1 DEM product, DEM coregistration needs to be carried out. For this purpose, the ICP (Iterative Closest Point) algorithm is used to align the Cartosat-1 DEM tile to the TanDEM-X DEM tile \cite{Ravanbakhsh2013}.

\section{TanDEM-X and Cartosat-1 DEM fusion}\label{sec:DEMfusion}

The results of relative and absolute accuracy assessment presented in \cite{Bagheri2017} demonstrated that over urban areas the height precision and resolution of the TanDEM-X DEM is not as fine as that of the Cartosat-1 DEM, whereas both DEMs have nearly identical quality over non-urban areas such as agricultural and forested fields. On the other hand, the TanDEM-X is absolutely more accurate than Cartosat-1 DEM. As explained in Section \ref{sec:Introduction}, data fusion techniques can be applied to take advantage of the properties of both kinds of DEMs to finally reach a DEM product with higher precision and absolute accuracy. 

While the simplest method for DEM fusion is weighted averaging, its main challenge is to employ suitable weight maps respective to each DEM. For example, the TanDEM-X and Cartosat-1 data can be fused by using weight maps delivered from the provided height error maps using simple weighted averaging:

\begin{equation}\label{WA}
\mathbf{D}_{F}=\mathbf{W}_{T}^{n}\odot\mathbf{D}_{T}+\mathbf{W}_{C}^{n}\odot\mathbf{D}_{C}
\end{equation}\

where, $ \mathbf{W}_{T} $  and $ \mathbf{W}_{C} $ are the normalized weights of the TanDEM-X and the Cartosat-1 heights, respectively, $ \mathbf{D}_{T} $ and $ \mathbf{D}_{C} $ are the height values taken from the TanDEM-X and Cartosat-1 DEMs, $ \odot $ denotes the element-wise product and $ \mathbf{D}_{F} $ refers to final fused DEM.

Usually, attached to InSAR DEMs such as TanDEM-X data, an additional product (called Height Error MAP: HEM) is provided, which roughly describes the quality of the generated DEM by the interferometric process based on coherence analysis \cite{Martone2012}, the number of the looks and the baseline configuration \cite{Just1994}. Similarly, For optical DEMs such as Cartosat-1 DEM, the quality map (also can be called HEM) is computed from stereo matching analysis.
Both HEMs are produced by error propagation analysis through the chain of DEM generation from the source data takes. 

The main disadvantage of HEM-based fusion is that HEMs are not always available in the form of DEM metadata, which holds in particular for DEMs generated from optical stereo data. This limits a broader applicability of HEM-based DEM fusion.   

For DEM fusion, pixel-wise weight maps should be created from the HEM maps. Two strategies can be pursued for weight map generation. The first one is to calculate weights as the inverse proportional of the squared height errors $e_{i}$:

\begin{equation}\label{WS1} 
w_{i}=\dfrac{1}{e_{i}^{2}}
\end{equation} 

Another way is to compute weights from the normalized residuals $e_{in}$:   

\begin{equation}\label{WS2} 
w_{i}=1-e_{in}
\end{equation}

\subsection{DEM fusion Support by Neural Network-Predicted weights}\label{subsec:ANN-WAfusion} 

\begin{figure}[ht!]
	\begin{center}
		\includegraphics[width=1\textwidth]{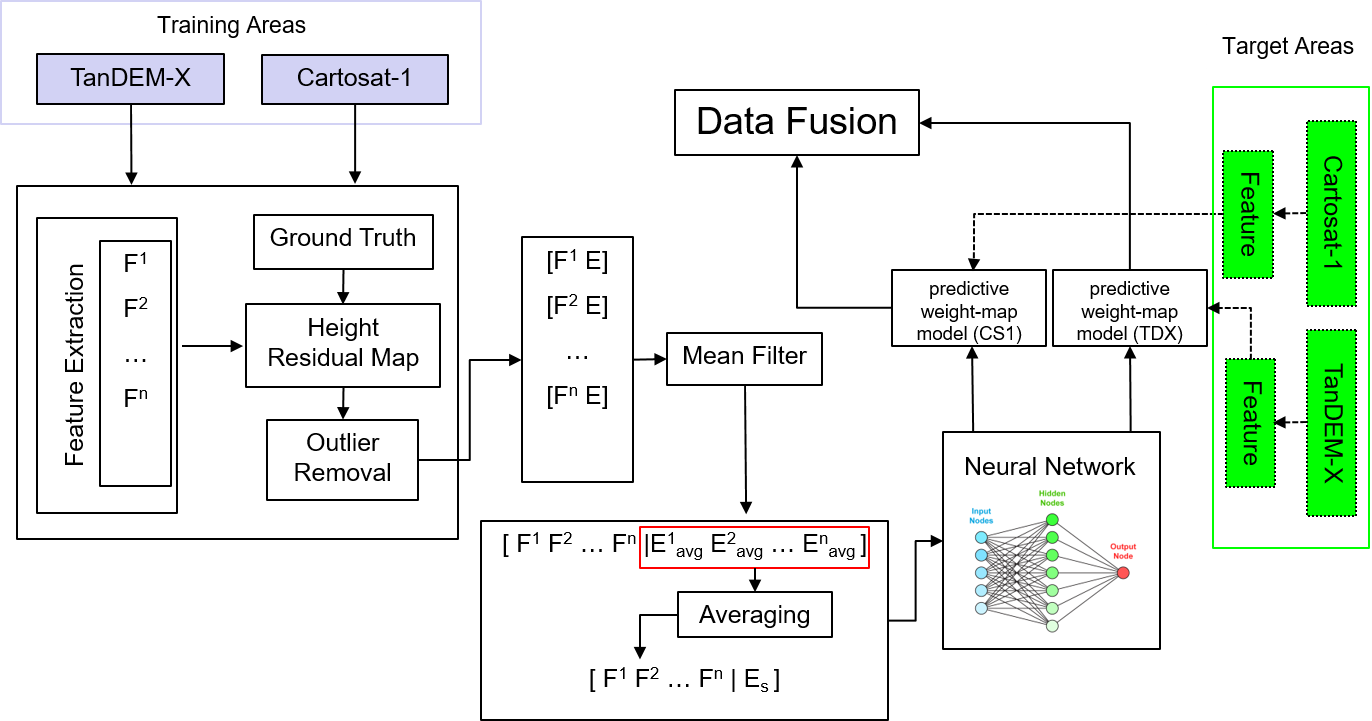}
		\caption{The framework designed to estimate the adaptive weights by ANN for the TanDEM-X and Cartosat-1 DEM fusion.}
		\label{framework}
	\end{center}
\end{figure}

As an alternative to the standard weight map generation process, in this paper we propose applying supervised learning that exploits high-resolution LiDAR ground truth data available for selected areas as training data to predict the height residual patterns and the corresponding weights for test data subsets. Fig \ref{framework} displays the framework of the proposed DEM fusion algorithm. In the heart of the proposed framework, an ANN is used to learn the relationship patterns of height errors and corresponding DEM features, which can subsequently be used for forecasting weight maps. The proposed framework (Fig \ref{framework}) can be summarized in three main steps:
\begin{enumerate}
	\item spatial feature extraction from DEMs and height error calculation
	\item data refinement 
	\item a) training the ANN on dedicated training subsets for which ground truth data is available and b) applying the ANN parameters to target subsets.  
	
\end{enumerate}
The output of the ANN is a predictive model that works as a weight predictor in target areas to which DEMs are fused based on the patterns explored in training subsets. More details of the framework's steps will be explained in the following subsections.

\subsubsection{Feature Extraction and Height Residual Computation}\label{subsec:feature}

For the training of the ANN, training data are selected from representatives of different land types that can usually be observed over urban areas. The description of these land types are in Section \ref{studyareades}. From those, different kinds of spatial features describing landscaping and roughness properties of the land surface are extracted. Several studies clarify the relationship between the spatial features and DEM qualities \cite{Toutin2002,4782702,Reinartz2010}. The following spatial features are useful for DEM fusion \cite{Olaya2009c}:
\begin{itemize}
   \item Geometrical parameters such as 1) Slope, which expresses the maximal rate of varying heights and 2) Aspect, which is the direction of the steepest slope in the mask window.  
   \item Statistical land surface parameters like 1) Anisotropic Coefficient of Variation (ACV), which describes the general geometry of local surfaces for distinguishing elongated and oval landforms; 2) Topographic Ruggedness Index (TRI), which is the 2D standard deviation filter; 3) Topographic Position Index (TPI), which is the difference between height of a pixel and mean height of neighboring pixels; 4) Roughness, which is  the largest height difference of target pixel and its surrounding cells; 5) Ruggedness, which is defined as the range value within an area; 6) Surface Roughness Factor (SRF), which is related to the normals to land surface; and 7) Entropy, which determines the uncertainty of height estimation in the selected window.  
\end{itemize} 

\begin{figure}[htbp]
	\centering
	
	\begin{subfigure}[b]{0.3\textwidth}
		\centering
		\includegraphics[width=1\textwidth]{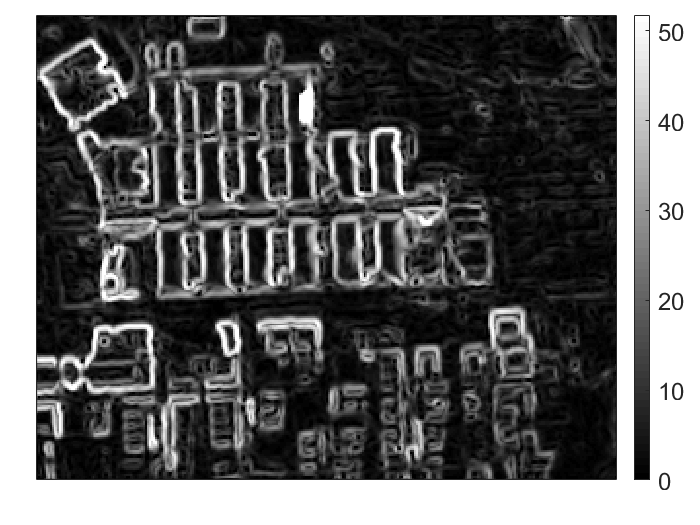}
		\caption{Slope}
		\label{Slope}
	\end{subfigure}
	\begin{subfigure}[b]{0.3\textwidth}
		\centering
		\includegraphics[width=1\textwidth]{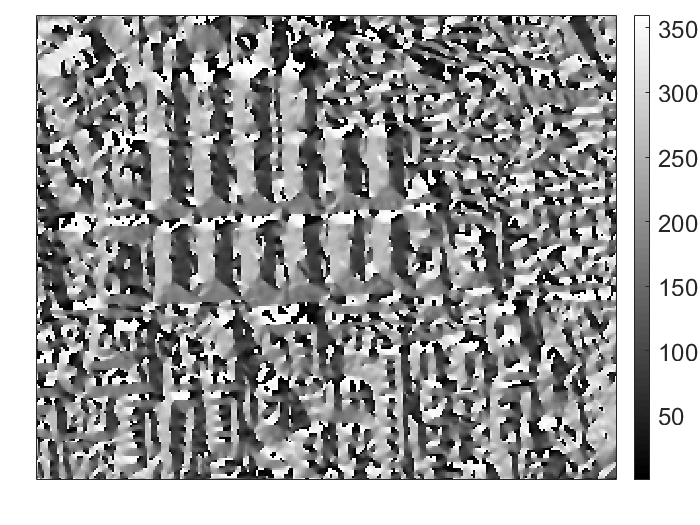}
		\caption{Aspect}
		\label{Aspect}
	\end{subfigure}    	
	\begin{subfigure}[b]{0.3\textwidth}
		\centering
		\includegraphics[width=1\textwidth]{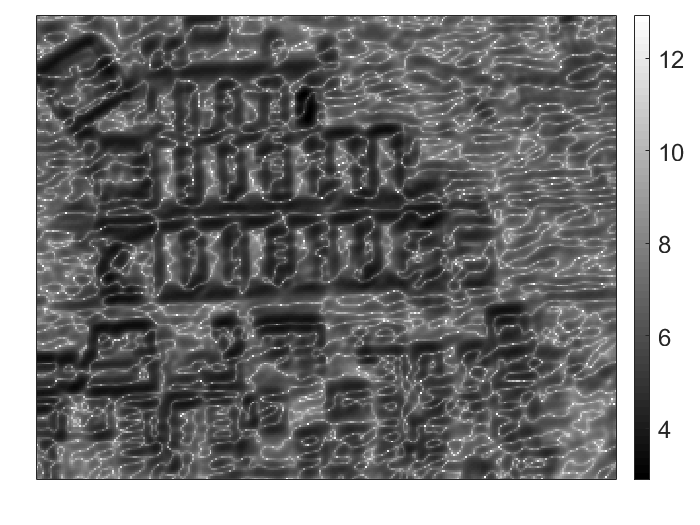}
		\caption{ACV}
		\label{ACV}
	\end{subfigure}
    \vspace{0.02\hsize}
    
	\begin{subfigure}[b]{0.3\textwidth}
		\centering
		\includegraphics[width=1\textwidth]{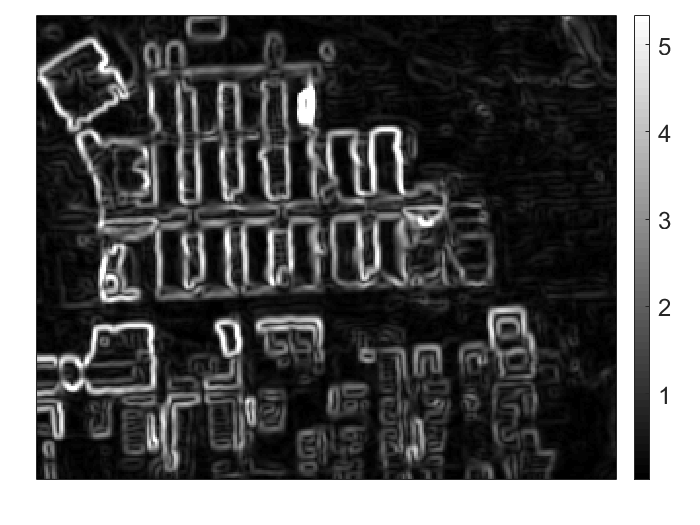}
		\caption{TRI}
		\label{TRI}
	\end{subfigure}
	\begin{subfigure}[b]{0.3\textwidth}
		\centering
		\includegraphics[width=1\textwidth]{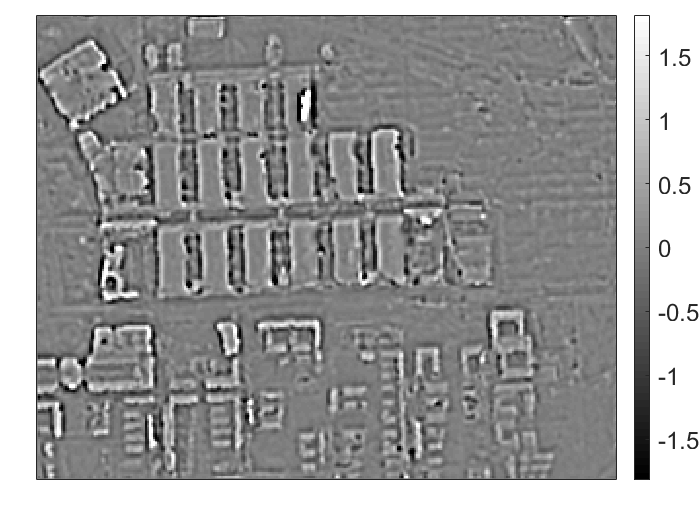}
		\caption{TPI}
		\label{TPI}
	\end{subfigure}	
	\begin{subfigure}[b]{0.3\textwidth}
		\centering
		\includegraphics[width=1\textwidth]{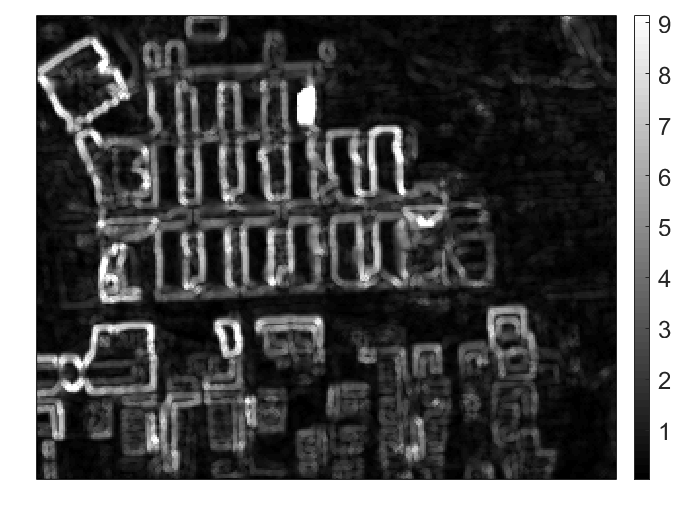}
		\caption{Roughness}
		\label{Roughness}
	\end{subfigure}
    \vspace{0.02\hsize}
    
	\begin{subfigure}[b]{0.3\textwidth}
	    \centering
	    \includegraphics[width=1\textwidth]{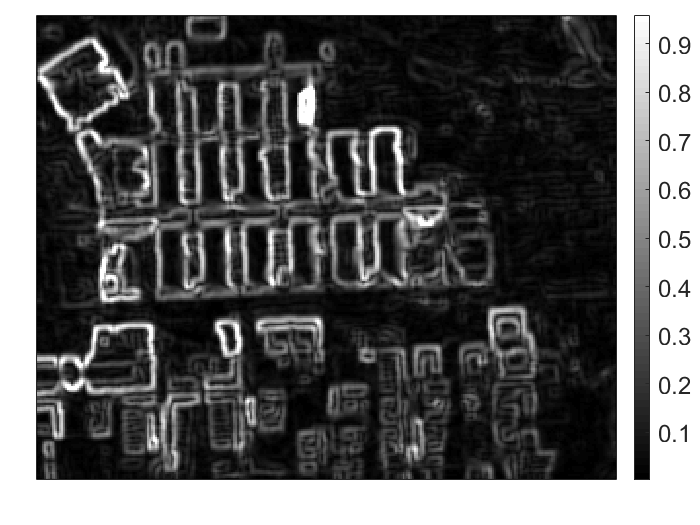}
	    \caption{Ruggedness}
	    \label{Ruggedness}
    \end{subfigure}
	\begin{subfigure}[b]{0.3\textwidth}
	    \centering
	    \includegraphics[width=1\textwidth]{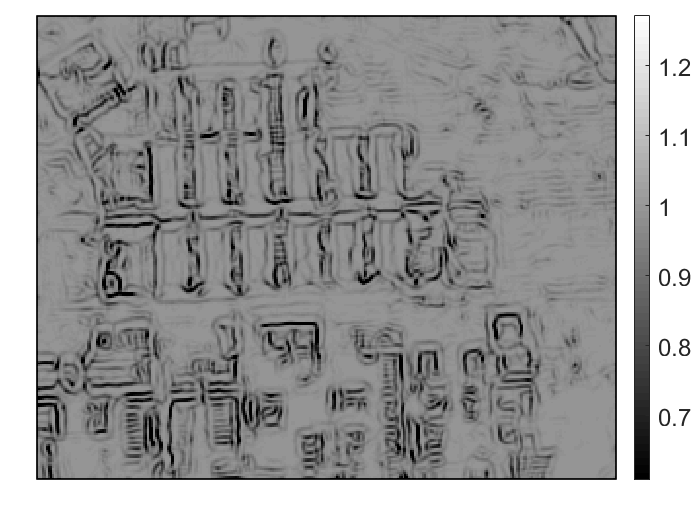}
	    \caption{SRF}
     	\label{SRF}
    \end{subfigure}
    \begin{subfigure}[b]{0.3\textwidth}
    \centering
    \includegraphics[width=1\textwidth]{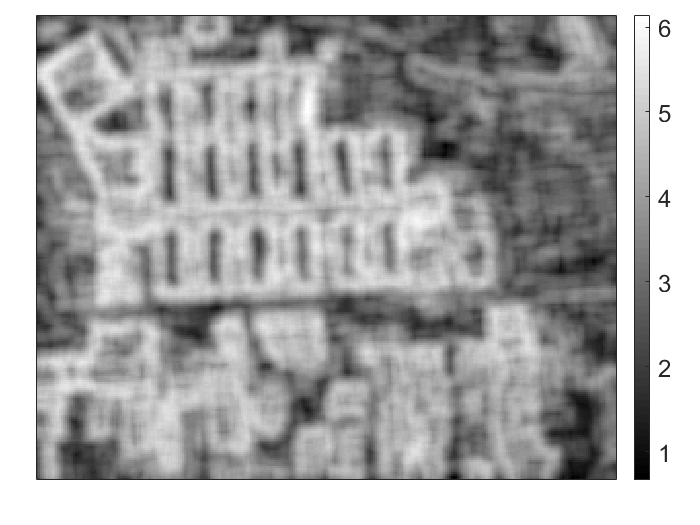}
    \caption{Entropy}
    \label{Entropy}
    \end{subfigure}
    \vspace{0.02\hsize}
    
    \begin{subfigure}[b]{0.3\textwidth}
	\centering
	\includegraphics[width=1\textwidth]{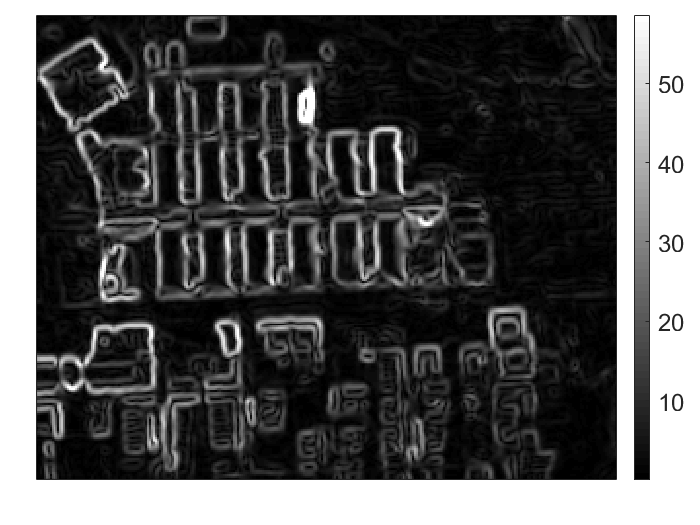}
	\caption{Edginess}
	\label{Edginess}
    \end{subfigure}
	\caption{Feature maps extracted from DEM from industrial area (I1) of Cartosat-1 DEM}
	\label{FeatureMap}
\end{figure}

 In addition to these parameters, edge values can also be extracted pixel-wisely by common edge detectors like the Sobel filter. At last, the HEM delivered with the TanDEM-X DEM and the quality map of Cartosat-1 DEM can also be used as a feature that reflects one source of induced errors in DEM. 
Apart from the HEM and the quality map, all mentioned features are extracted by convolution with a 3$\times$3 square window as a mask around each cell. Fig \ref{FeatureMap} exemplarily shows the maps for  these features extracted from the Cartosat-1 data in the industrial area, subset I1.

Moreover, height residual maps are calculated for all training subsets by subtracting the LiDAR ground truth elevations from the corresponding DEM elevations.
The obtained feature maps, along with these height residuals, are used to train an ANN to model the relationship between feature values and height errors. 
Fig \ref{FeatureError} exemplarily depicts the variation of errors with respect to feature values in the subset of the industrial area (I1) for both study DEMs. The height residuals, at first, are very noisy, so that a low pass filter (details explained in the following subsection) is required to reveal the error patterns as depicted in Fig \ref{FeatureError}.
 
\subsubsection{Data Refinement}\label{subsec:Preprocess}

Prior to constituting the ANN structure, another important step is to refine the height errors oriented to extracted spatial feature values to get rid of outliers and decrease the noise effects. The calculated height residuals are polluted by high-frequency noise, which will affect the training of the ANN.

\begin{figure} 
	\centering
	\begin{subfigure}[b]{1\textwidth}
		\centering
		\includegraphics[width=0.7\textwidth]{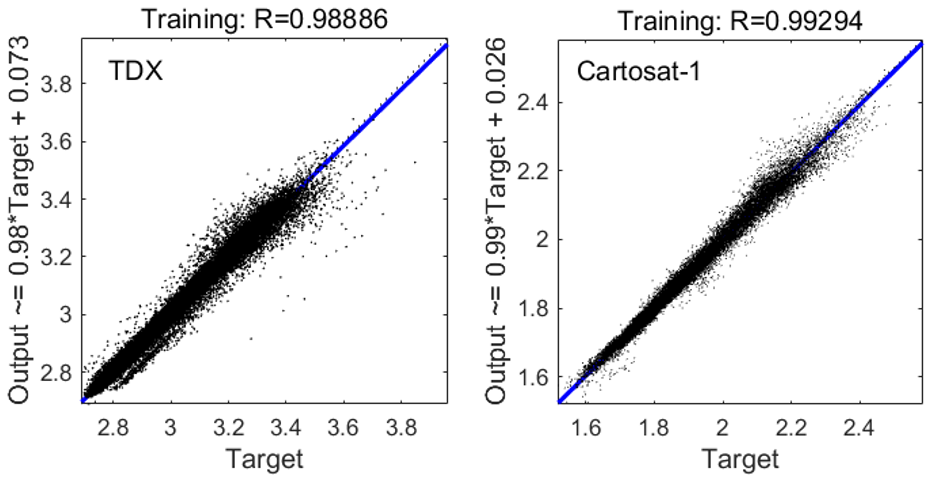}
		\caption{Results after carrying out the proposed refinement steps.}
		\label{Reg1}
	\end{subfigure}
	\hfil
	
	\begin{subfigure}[b]{1\textwidth}
		\centering
		\includegraphics[width=0.7\textwidth]{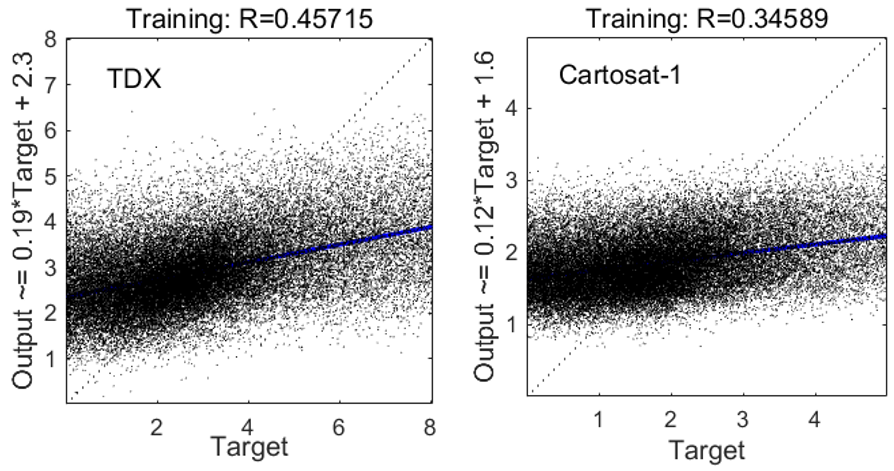}
		\caption{Results without using the proposed refinement.}
		\label{Reg2}
	\end{subfigure}
	\caption{Regression plot of training data for industrial area (Sub D1)}
	\label{Regression}
\end{figure}

The performance of the network in the case of using smoothed residual maps derived from the refinement step, as well as using raw data without smoothing, only removing the outliers, are illustrated in Fig \ref{Regression}.
Fig \ref{Reg2} indicates that noisy height residuals disrupt the training procedure and prevent the ANN from recognizing the error patterns. The correlation evaluation of desired outputs and results achieved by ANNs demonstrates the efficiency of the refinement step. By employing the refinement framework, the networks can learn the error pattern respective to features with high correlation (more than 0.98) between outputs of training and target values for TanDEM-X and Cartosat-1 DEMs, respectively. Without implementing the refinement, the training performances of networks are significantly lower, illustrated by an output with correlations lower than 0.50 and 0.35 for the TanDEM-X and the Cartosat-1 DEM, respectively. 

To reduce the noise effects with the aim of promoting the training, a smoothing process, characterized by two-step mean filtering is carried out. 
The first step of the refinement is to bin the feature values that can be obtained by a simple empirical-statistical binning technique. The feature values $f_{i}^{j}$ correspond to the height residuals $e_{i}$ by:

	\begin{equation}\label{FE1}
	\left[ \begin{array}{cccc}
	f_{1}^{1} &f_{1}^{2}&...&f_{1}^{n} \\
	f_{2}^{1} &f_{2}^{2}&...&f_{2}^{n}\\
	... &...&...&...\\
	f_{m}^{1} &f_{m}^{2}&...&f_{m}^{n}\\
	\end{array}\right] \Leftrightarrow 	\left[ \begin{array}{cccc}
	e_{1}  \\
	e_{2} \\
	... \\
	e_{m} \\
	\end{array}\right]
	\end{equation}
    
where $f_{i}^{j}$ is the value of the feature $j\in\{1,2,...,n\}$ and $e_{i}$ is corresponding height residual value in pixel $i\in\{1,2,...,m\}$.
	
At first, errors exceeding 3$\times$NMAD ($e_{\zeta}$, which are identified by index $\zeta$) are detected as outliers and then eliminated along with their corresponding feature values $\left[\begin{array}{cccc} f_{\zeta}^{1} &f_{\zeta}^{2}&...&f_{\zeta}^{n}\end{array}\right]$ from the training dataset. The normalized median absolute deviation (NMAD) is recommended as a robust accuracy measure rather than the classical root mean square error (RMSE) for the mitigation of outliers affecting the elevation data of the study DEMs \cite{Hoehle2009}. 

The relation (\ref{FE1}) can be rewritten in the form of feature vectors that include the values of each feature type for all pixels of the DEM:

    \begin{equation}\label{FE2}
     \begin{aligned}    
    \left[ \begin{array}{cccc}
    \mathbf{F^{1}} &\mathbf{F^{2}}&...&\mathbf{F^{n}}
    \end{array}\right] \Leftrightarrow 	
    \mathbf{E}\\    
     \mathbf{F^{j}}=	\left[ \begin{array}{cccc}
     f_{1}^{j} &f_{2}^{j}&...&f_{m}^{j}\end{array}\right]^{T}
     \end{aligned}
    \end{equation}

\begin{figure}[htbp]
	\centering
	
	\begin{subfigure}[b]{0.3\textwidth}
		\centering
		\includegraphics[width=1\textwidth]{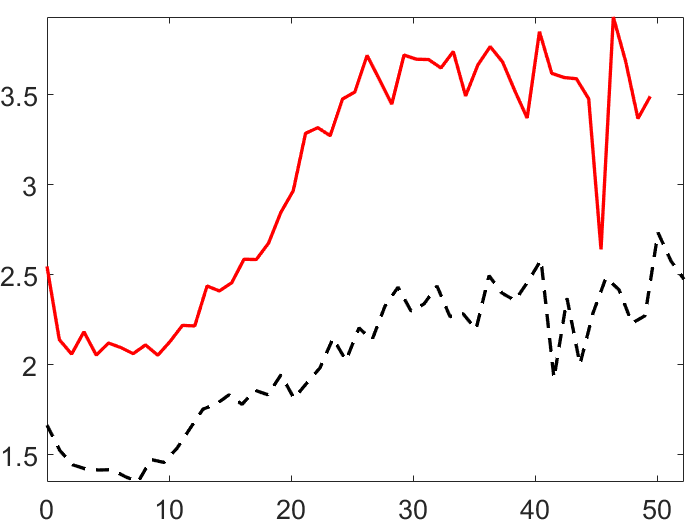}
		\caption{Slope}
		\label{SlopeError}
	\end{subfigure}
	\begin{subfigure}[b]{0.3\textwidth}
		\centering
		\includegraphics[width=1\textwidth]{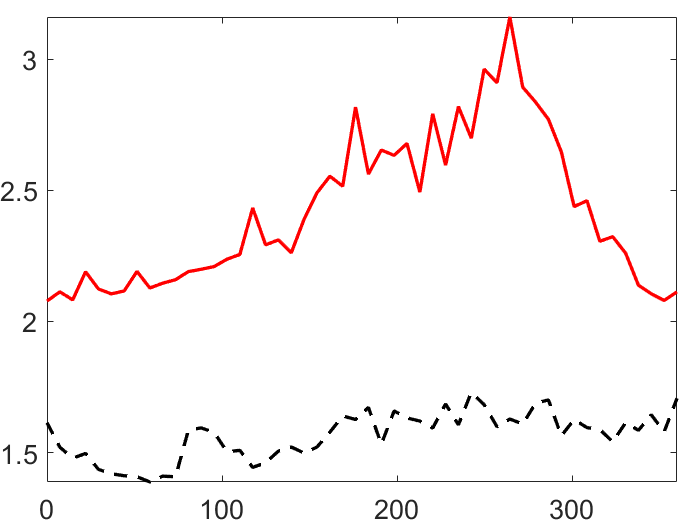}
		\caption{Aspect}
		\label{AspectError}
	\end{subfigure}    	
	\begin{subfigure}[b]{0.3\textwidth}
		\centering
		\includegraphics[width=1\textwidth]{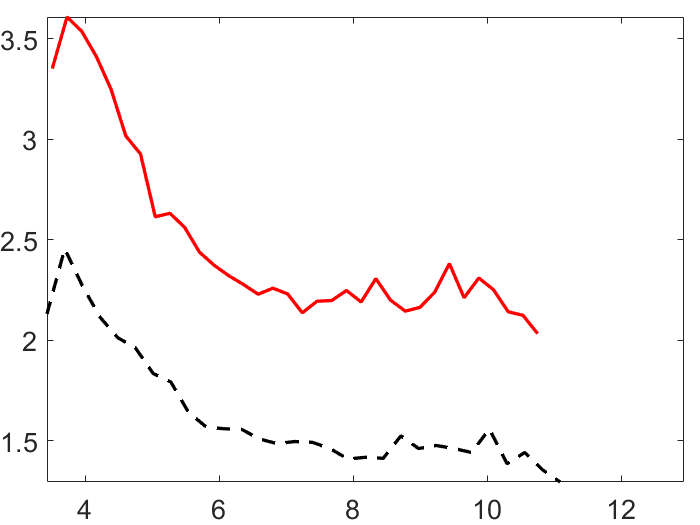}
		\caption{ACV}
		\label{ACVError}
	\end{subfigure}
	\vspace{0.02\hsize}
	
	\begin{subfigure}[b]{0.3\textwidth}
		\centering
		\includegraphics[width=1\textwidth]{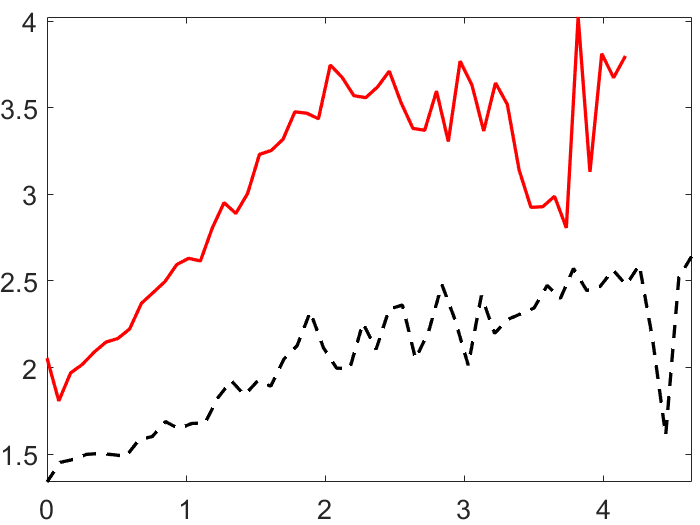}
		\caption{TRI}
		\label{TRIError}
	\end{subfigure}
	\begin{subfigure}[b]{0.3\textwidth}
		\centering
		\includegraphics[width=1\textwidth]{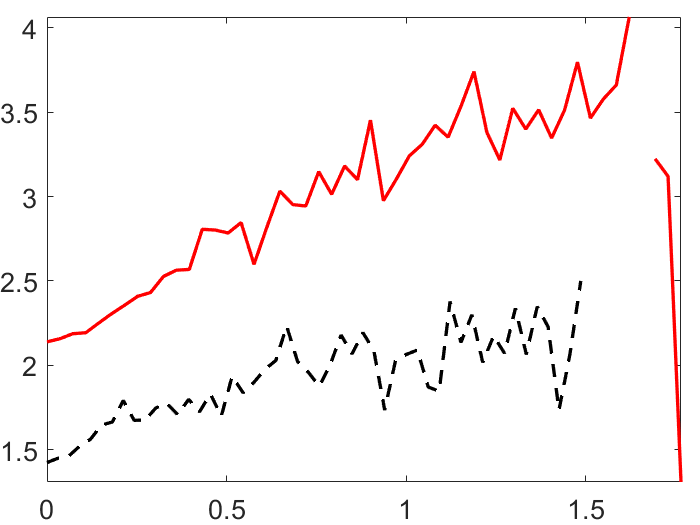}
		\caption{TPI}
		\label{TPIError}
	\end{subfigure}	
	\begin{subfigure}[b]{0.3\textwidth}
		\centering
		\includegraphics[width=1\textwidth]{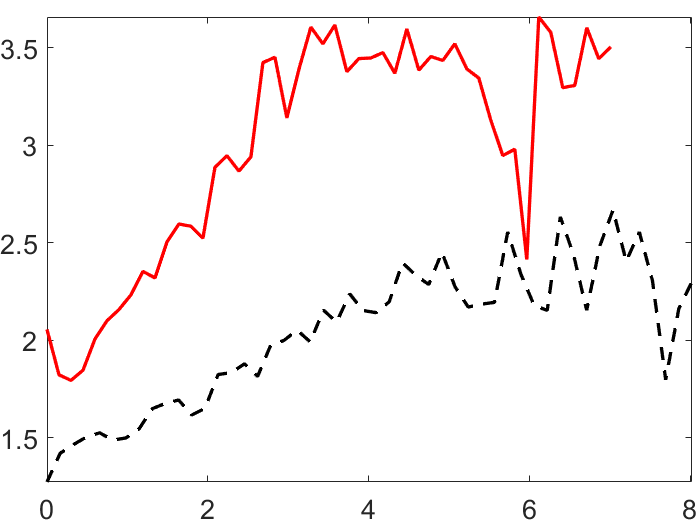}
		\caption{Roughness}
		\label{RoughError}
	\end{subfigure}
	\vspace{0.02\hsize}
	
	\begin{subfigure}[b]{0.3\textwidth}
		\centering
		\includegraphics[width=1\textwidth]{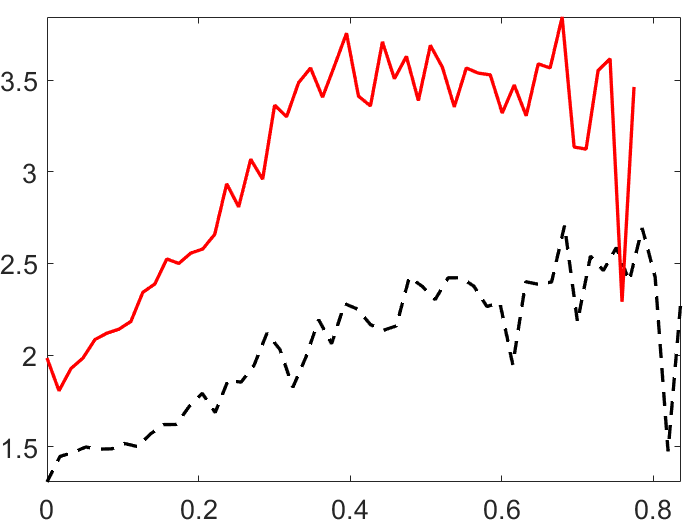}
		\caption{Ruggedness}
		\label{RugError}
	\end{subfigure}
	\begin{subfigure}[b]{0.3\textwidth}
		\centering
		\includegraphics[width=1\textwidth]{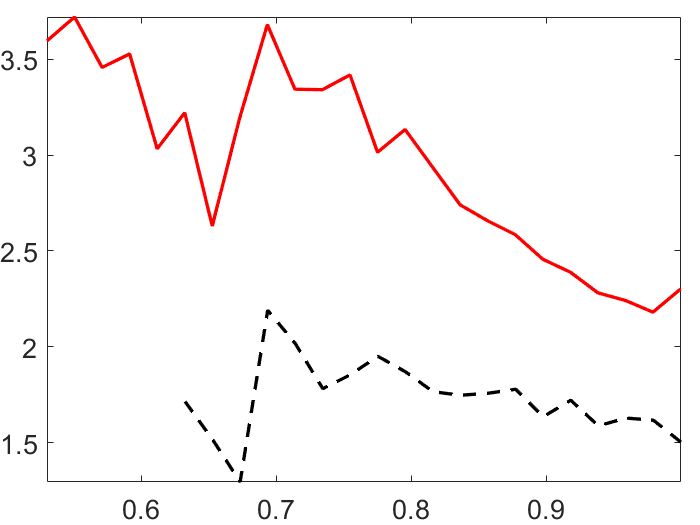}
		\caption{SRF}
		\label{SRFError}
	\end{subfigure}
	\begin{subfigure}[b]{0.3\textwidth}
		\centering
		\includegraphics[width=1\textwidth]{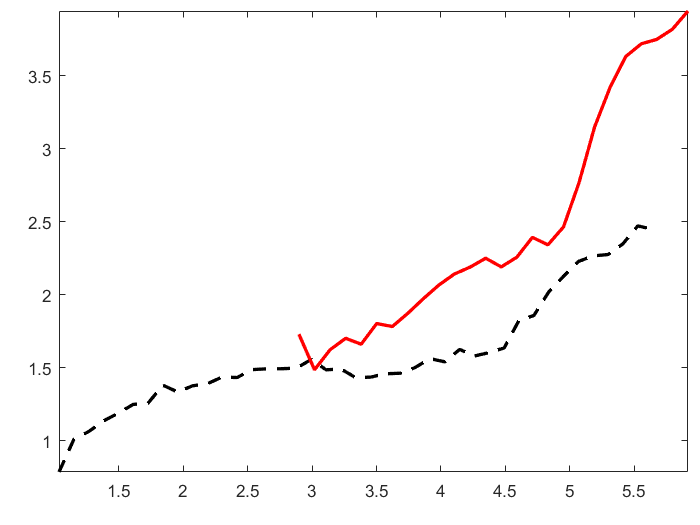}
		\caption{Entropy}
		\label{EntError}
	\end{subfigure}
	\vspace{0.02\hsize}
	
	\begin{subfigure}[b]{0.3\textwidth}
		\centering
		\includegraphics[width=1\textwidth]{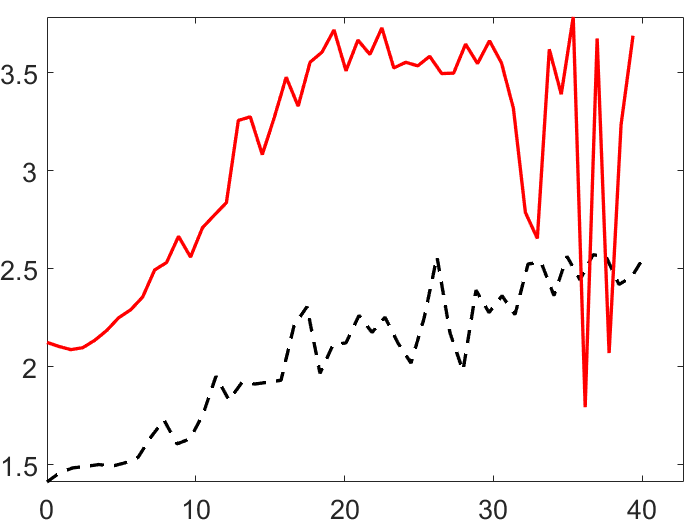}
		\caption{Edginess}
		\label{Edgerror}
	\end{subfigure}
	\caption{Height error patterns of the TanDEM-X DEM (in red) and the Cartosat-1 DEM (in black dashed) for industrial area (Subset I1): horizontal directions show the feature values and vertical directions indicate mean absolute height residuals in each bin achieved from the refinement step.}
    \label{FeatureError}
    \end{figure}

After removing outliers, The values of the feature vector ($\mathbf{F^{j}}$) and their corresponding height residuals $ \mathbf{E} $ are binned by the Freedman-Diaconis rule \cite{Birge2006}:

\begin{equation}\label{bin}
N=\frac{f_{max}^{j}-f_{min}^{j}}{h}
\end{equation}

where $h = 2\times I\times k^{-1/3} $. The output of above formulation is the number of bins ($N$) for feature $j$ with bin width of $h$, just by detecting the max and min values of  measured feature ($f_{max}^{j}$ and $f_{min}^{j}$). $I$ is the interquartile range and $k$ is the number of measurements that are remaining height residuals after outlier removal. In other words, $ k $ refers to number of pixels whose height errors ($e_{i}$) are lower than the threshold 3$\times$NMAD. The mean filter is applied bin-wise to generate smoother height residual. The output of this filtering is the numerical feature-error model in which each feature vector $\mathbf{F^{j}}$ corresponds to a new smoothed height residual map $\mathbf{E_{avg}^{j}}=\left[ \begin{array}{cccc}e_{1avg}^{j} &e_{2avg}^{j}&...&e_{mavg}^{j}\end{array}\right]^{T}$. It has to be noted that infrequent feature values are thrown away  by a threshold.
This procedure should be followed for each type of feature. Fig \ref{FeatureError} presents the graphical depictions of feature-error models derived for an industrial area (subset I1) for TanDEM-X and Cartosat-1 after binning and mean filtering. Consequently, for each pixel, n height residual values at last are acquired. This means there are n residual maps, which are linked to n numerical feature-error models ($\left[\begin{array}{ccccccccc} 
\mathbf{F^{1}} &\mathbf{F^{2}}&...&\mathbf{F^{n}}&| &\mathbf{E_{avg}^{1}} &\mathbf{E_{avg}^{2}}&...&\mathbf{E_{avg}^{n}} \end{array}\right]$). 

Next, the second step of the smoothing process is to average again the achievements of the former step (smoothed height residuals) to finally create a unique height residual.  After this refinement, the data are ready to insert into the ANN for training and exploring the patterns.

\subsubsection{Weight Map Generation by ANN}\label{subsec:ANNWeightMap} 
\begin{figure}[ht!]
 	\begin{center}
 		\includegraphics[width=1\textwidth]{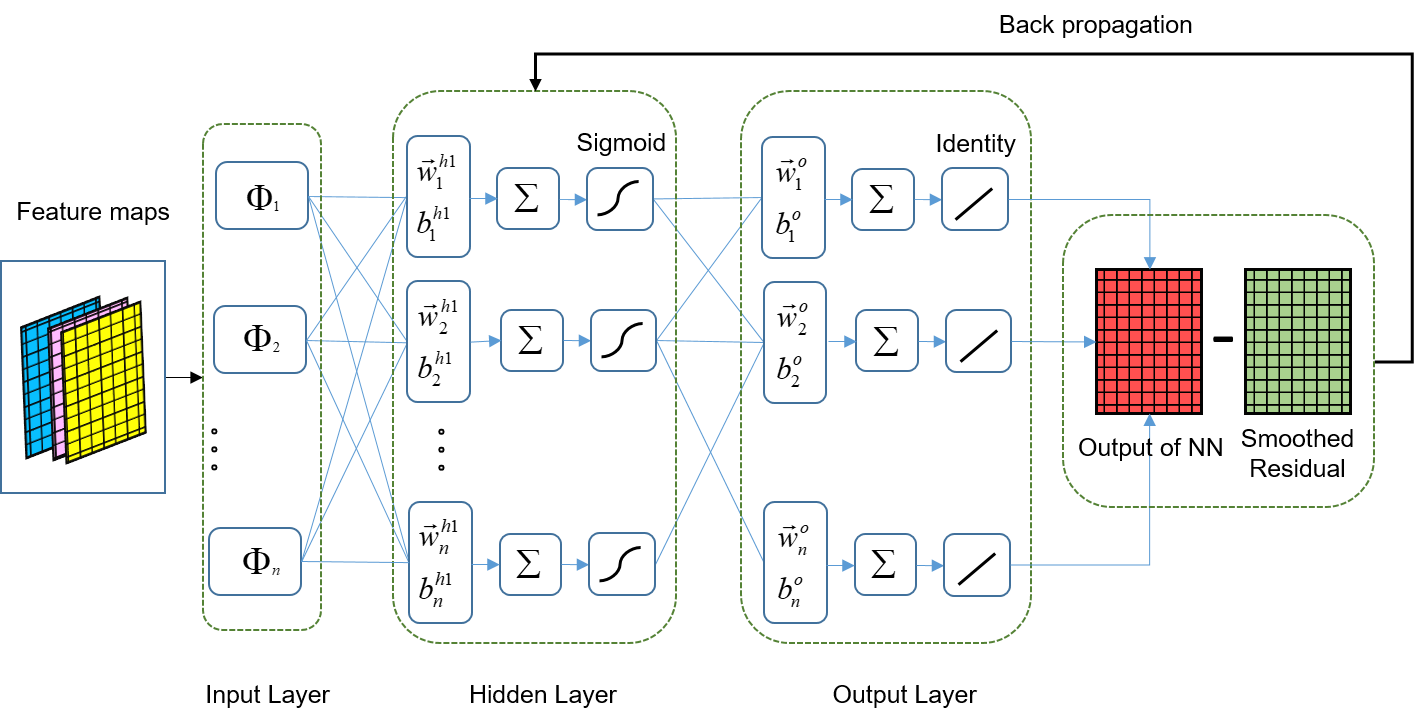}
 		\caption{Structure of neural network for weight map prediction }
 		\label{NN}
 	\end{center}
 \end{figure}

The filtered outcomes from the previous stage are employed to train a fully connected feed-forward neural network.  After feature extraction, height residual computation and refinement based on the pipeline described in Section \ref{subsec:Preprocess}, the outputs become the input into the ANN.  The ANN is trained using the filtered feature vectors as inputs and the modified height residuals as outputs, which are cast in the form of:

 \begin{equation}\label{finalFE}  
 \begin{aligned}  
 \left[\begin{array}{cccccc} 
 \mathbf{\Phi_{1}} &\mathbf{\Phi_{2}}&...&\mathbf{\Phi_{m}}&| &\mathbf{E_{s}} \end{array}\right], \\
  \textrm{where} \quad \mathbf{\Phi_{i}}=\left[\begin{array}{cccc} 
 f_{i}^{1} &f_{i}^{2}&...&f_{i}^{n} \end{array}\right]^{T}
 , \quad i\in\{1,2,...,m\}
 \end{aligned} 	
 \end{equation}
 
contains the values of the different features for a given pixel $i$ and $E_{s}$ is the final smoothed residual map obtained through the two-step mean filtering. Fig \ref{NN} shows the structure of the network, which consists of an input layer in which neurons with the label of the feature values of each pixel ($\mathbf{\Phi_{i}}$) are connected to the smoothed height residual of the corresponding pixel through the hidden layers. In the repetitive process, with back propagation training, the weights of neurons are gradually modified to decrease the discrepancy between smoothed height residual maps and the map achieved by the network. The main achievement of the ANN after successful training is that a model can estimate the weight maps for each part of DEMs from forecasting the height residuals just by measuring the spatial features. 

\begin{figure}[ht!]
	\begin{subfigure}[b]{0.5\textwidth}
		\centering
		\includegraphics[width=0.95\textwidth]{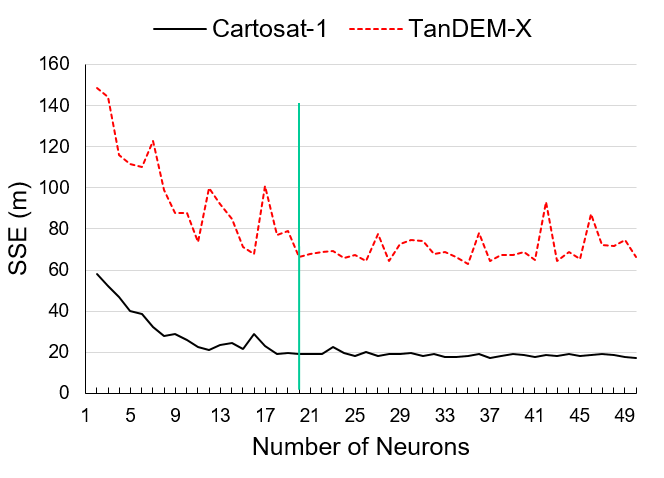}
		\caption{}
		\label{SSE1}
	\end{subfigure}
	\hfil 	
	\begin{subfigure}[b]{0.5\textwidth}
		\centering
		\includegraphics[width=0.95\textwidth]{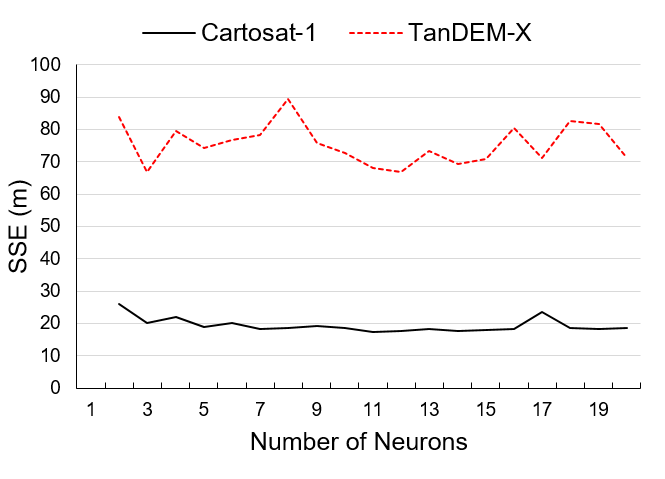}
		\caption{}
		\label{SSE2}
	\end{subfigure}
	\caption{The performance of the ANNs with different structures measured by SSE (Sum of Squares Error); a) The structure with one hidden layer. b) The structure that organized by two hidden layers: first, with number of neurons fixed to n=20, and second with varying number of neurons}
	\label{NN-SSE1}
\end{figure}

The optimal structures of ANNs for both study DEMs were investigated by tracking the cost function values during the training stage. The performances of networks were evaluated by considering one hidden layer and changing the number of neurons in this layer. Then, deeper networks were examined by adding another hidden layer. Figures \ref{SSE1} and \ref{SSE2} depict the performance of the neural networks on test data for an increasing number of neurons in the ANNs with one hidden layer, as well as for structures designed with two hidden layers. The plots display that the rise in the number of neurons in the first layer is more influential than adding the layers and making the networks deeper. In ANNs with one hidden layer, the general trend of changing costs remained stable after adding more than 20 neurons. In other words, adding more neurons to a hidden layer does not change the performance of the networks and the subsequent DEM fusion result. This evaluation systematically determined the optimal structure of NNs considered for both DEMs.

\section{TanDEM-X and Cartosat-1 DEM Fusion Results}\label{sec:fusionresult}

One additional advantage of Cartosat-1 and TanDEM-X DEM fusion is to increase the absolute vertical accuracy with respect to the data accuracy provided by the original Cartosat-1 DEM. By successful alignment of the Catosat-1 and TanDEM-X DEMs, the Cartosat-1 DEM is vertically positioned in the location of TanDEM-X, which indicates an improvement of the absolute geolocation through DEM fusion. 
After vertical alignment, the HEM- and ANN-based approaches are examined to increase the height precision of both DEMs.

HEM-based fusion was implemented for all subsets except subsets F1 and L, due to unavailability of values of STD of matching for these areas. The HEM-based fusion results for the other subsets are presented in Tab \ref{HEMresult}. The common metric for measuring the accuracy of DEMs is root mean square error (RMSE). Additionally, Normal Median Absolute Deviation (NMAD) is used as another measure for height precision analysis \cite{Hoehle2009}. The RMSE measure reflects the effects of whole larger and small errors but NMAD implies the presence of small errors in elevation data.        

\begin{sidewaystable}
	\centering
	\begin{tabular}{llllllllll}\hline
		\multicolumn{2}{c}{Areas}&\multicolumn{2}{c}{TanDEM-X}&\multicolumn{2}{c}{Cartosat-1}&\multicolumn{2}{c}{Fused DEM}\\\hline
		
		&&NMAD&RMSE&NMAD&RMSE&NMAD&RMSE\\
		\multirow{8}{*}{Urban} 
		&Industrial: D1	    &2.68 & 4.61 &\cellcolor{green!20}1.66  &\cellcolor{green!20}3.24  & 1.95 & 3.57    \\ 
		&Industrial: D2	    &3.44 & 5.12 &\cellcolor{green!20}2.44  &\cellcolor{green!20}3.56  & 2.92 & 4.27    \\ 
		&Inner city: I1	    &5.57 & 6.43 &\cellcolor{green!20}4.34 &\cellcolor{green!20}5.27  & 4.64 & 5.33    \\ 
		&Inner city: I2	    &5.92 & 5.81 &\cellcolor{green!20}4.63  &5.22  & 5.09 & \cellcolor{green!20}5.13    \\ 
		&High building: H1	&6.72 & 18.17&\cellcolor{green!20}3.63 &\cellcolor{green!20}13.10  & 4.69 & 16.12  \\ 
		&High building: H2	&3.76 & 8.41 &3.29  &8.13  &\cellcolor{green!20}3.14 &\cellcolor{green!20}7.93   \\ 
		&Residential: R1	&2.95 & 3.10 &\cellcolor{green!20}2.56  &\cellcolor{green!20}2.83  & 2.81 & 2.90    \\ 
		&Residential: R2	&2.30 & 2.61 &\cellcolor{green!20}2.02  &2.45  & 2.13 & \cellcolor{green!20}2.44    \\ 
		
		\multirow{4}{*}{Non-Urban}    
		&Forested: F1     	&---   &  ---  &---    &   --- &---    & ---    \\ 
		&Forested: F2     	&3.88 & 4.82 &\cellcolor{green!20}3.23  & 4.65 &3.51  &\cellcolor{green!20} 4.35  \\ 
		&Agricultural (F1)	&---   &  ---  &---    &  ---  &---    & ---    \\ 
		&Agricultural (A)	&0.93 & 0.84 &\cellcolor{green!20}0.65  & 0.81 &0.98  & \cellcolor{green!20}0.76  \\ 
		\hline	
			
   \end{tabular}
    \caption{Height precision (in meters) of TanDEM-X DEM, the Cartosat-1 DEM and final fused DEM using the HEMs and STD of matching as weight maps in WA (in meters). The green shaded values indicate the best values of metrics relevant to the quality of the compared DEMs}
    \label{HEMresult}
\end{sidewaystable}

The results indicate that using the HEM of the TanDEM-X data and STD of matching for Cartosat-1 DEM could only increase the height precision of TanDEM-X elevations and an improvement for Cartosat-1 data cannot always be fulfilled for all land types.

In ANN-based fusion approach, training data are selected from diverse land types mentioned in Tab \ref{studyareades}, such as industrial, inner city, residential areas, high building subset, agricultural and forested areas. The usage of training data from different land types guarantees the presence of all possible values of features respective to height residuals in the process of pattern recognition by the NN, and give the assurance of discovering a more general model that can be used for any arbitrary land type. After successful training, the ANN can be applied for predicting the height residual in selected target areas where two DEMs are supposed to be fused. The predicted residual maps are used as weight maps in the weighted average fusion. Thus, two separate ANNs are needed for both the TanDEM-X and Cartosat-1 DEM  to generate individual weight maps for each DEM separately. For the experiments in this paper, two strategies are followed regarding the combination of training subsets. 

In strategy A, separate ANN are trained for each specific land type in order to  provide class-specific weight map predictions. In this case, the subsets D1, I1, H1, R1 as well as F1 separated into its forested and its agricultural segments are used as training subsets, on which six different ANNs are trained. The resulting weight map predictors that are used for DEM fusion in the respective target areas are then applied on the corresponding test areas D2, I2, H2, R2, F2 and A.

In strategy B, all subsets of all different land types are simultaneously used as training data to create a general predictor model that can be used for weight map generation in arbitrary target subsets. In this experiment, the subsets D1, I1, H1, R1 and F1 are used to train the ANN, and the resulting output model is used for all target subsets. 

In both experiments, 70\% of data from the training subsets are devoted to training, and 15\% are for validation to control the training process in order to avoid over-fitting and under-fitting. However, the whole process of the proposed framework will be implemented on the independent subsets, tuning the networks' parameters such as depth and number of neurons in each layer, and the remaining data (15\%) are devoted to monitoring the performance of NN during the training.

Tab \ref{weighting} presents the results of DEM fusion over individual target areas, employing different strategies of data selection for training. The results show nearly identical results for both strategies. Thus, pouring all subsets from different land types to make a general predictor model decreases the number of necessary ANNs from six to one for each kind of DEM. On the contrary, the size of the input data for the training becomes larger; thus, training requires more runtime. The runtime of NN training adopting different strategies (A and B), implemented in a system equipped with  Intel(R) Core(TM) i7-6700, 3.40 GHz CPU and 16 GB RAM is collected in Tab. \ref{runtime}. The total runtime of strategy A for training both networks of Cartosat-1 and TanDEM-X is 158.1 seconds while training according to strategy B takes 519.6 seconds. This confirms that strategy A is computationally more economical for training.     

\begin{table*}
	\centering
	\begin{tabular}{ccll}\hline
		Strategy&DEM &Subset& \makecell{Runtime \\(second)}\\\hline	
		\multirow{12}{*}{A} 
		&\multirow{6}{*}{Cartosat-1}
		&Industrial: D1	&33.2	    \\ 
		&&Inner city: I1	&10.8	   \\ 
		&&High building: H1	&0.1	    \\ 
		&&Residential: R1	&0.8	    \\ 
		&&Forested: F1	    &2.4	   \\ 
		&&Agricultural: F1	&8.9	  \\\cline{2-4}
		& \multirow{6}{*}{TanDEM-X}
		&Industrial: D1	&59.7	    \\ 
		&&Inner city: I1	&15.6	   \\ 
		&&High building: H1	&0.1	    \\ 
		&&Residential: R1	&1.4	    \\ 
		&&Forested: F1	    &6.8	   \\ 
		&&Agricultural: F1	&18.3	  \\\hline\hline
		\multirow{2}{*}{B}  
		&Cartosat-1	&All	&187.5	    \\
		&TanDEM-X	&All	&332.1	    \\\hline	
	\end{tabular}
	\caption{The training computational cost using different strategies, A and B}
	\label{runtime}
\end{table*}

In Tab \ref{weighting}, we can also observe the RMSEs of fused DEMs generated by the weight maps that computed by adopting different weighting strategies. The results display slightly lower RMSE values by the $\dfrac{1}{e_{i}^2}$ weighting formulation for some areas like industrial, inner city and forested areas. On the other, the qualities of fused DEMs following the different training strategies (A and B) are almost same. The benefit of using strategy B is the establishment of a general predictor model that can be used for any arbitrary land types; while in strategy A, the networks have to be trained six times (for six different land types) and achieved predictor models must be used for respective land types in target areas, requiring semantic classification of the study area to identify different land types for DEM fusion.

\begin{table*}
	\centering
	\begin{tabular}{lllll}\hline
		\multicolumn{1}{c}{}&\multicolumn{4}{c}{Fused DEM}\\\hline
		Training strategy&Individual&Individual&All&All\\
		Weight& $\dfrac{1}{e_{i}^{2}}$& $1-e_{in}$& $\dfrac{1}{e_{i}^{2}}$& $1-e_{in}$\\
		Areas&RMSE&RMSE&RMSE&RMSE\\
		Industrial: D2	    &3.46 &3.63 &3.52& 3.62     \\ 
		Inner city: I2	    &4.84 &4.94 &4.85&4.92      \\ 
		High building: H2	&7.75 &7.71 &7.75& 7.66    \\ 
		Residential: R2	&2.33 &2.38 &2.34&  2.35    \\ 
		Forested: F2     	&4.18 &4.28 &4.18& 4.25     \\ 
		Agricultural: A	&0.70 &0.69 &0.68&0.69      \\\hline	
	\end{tabular}
	\caption{Results of fusing TanDEM-X and Cartosat-1 DEM (in meters) by employing weight maps predicted by ANN with the application of different training strategies and different types of weighting.}
	\label{weighting}
\end{table*}

A final experiment was implemented for comparing the DEM fusion results by employing ANN-based and HEM-based approaches with the initial DEMs. For this purpose, using ANNs with one hidden layer and 20 neurons in the hidden layer and the training based on the subsets D1, I1, H1, R1 and F1 (following strategy B as the economical way), the predictive models were created.
At last, the full DEM fusion chain was carried out on totally independent subsets to evaluate the  capability of the proposed algorithm for DEM fusion. These independent subsets were selected as target areas from different land types (subsets D2, I2, H2, R2, F2 and A). The TanDEM-X and Cartosat-1 DEMs of these subsets were fused together by weighted averaging using the achieved height error maps as weights of the input DEMs by the $\dfrac{1}{e_{i}^2}$ weighting formulation. It has to be noted, as illustrated in Section \ref{sec:Introduction}, the quality of the TanDEM-X DEM is significantly worse than the Cartosat-1 DEM in the Lake subset, so that heights of the TanDEM-X DEM should be completely substituted by the heights of the Cartosat-1 DEM.  

\begin{figure}[htb]
	\begin{minipage}[b]{0.48\linewidth}
		\centering
		\centerline{\epsfig{figure=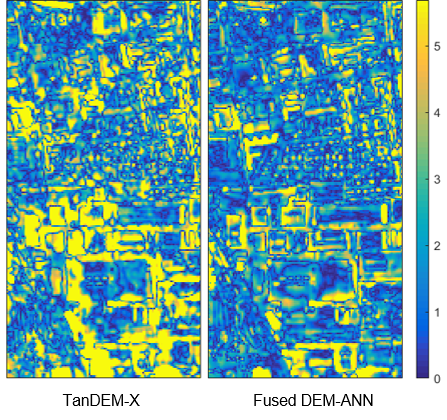,width=1\columnwidth}}
		\vspace{0.05cm}
		\centerline{Sub D2}\medskip
	\end{minipage}
   	\begin{minipage}[b]{0.48\linewidth}
   		\centering
   		\centerline{\epsfig{figure=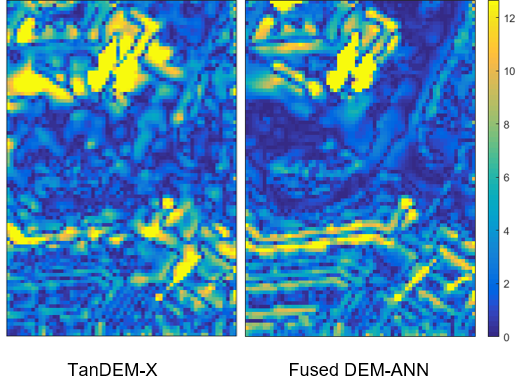,width=1\columnwidth}}
   		\vspace{0.05cm}
   		\centerline{Sub H2}\medskip
   	\end{minipage}
   
   	\begin{minipage}[b]{1\linewidth}
   	\centering
   	\centerline{\epsfig{figure=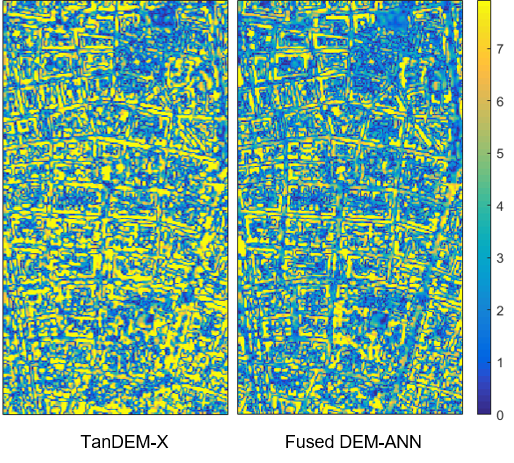,width=0.48\columnwidth}}
   	\vspace{0.05cm}
   	\centerline{Sub I2}\medskip
   \end{minipage}

	\caption{Absolute residual maps of TanDEM-X and Fused DEM using ANN-predicted weight maps in some exemplary subsets.}
	\label{Residualfusion}
\end{figure}

The results are summarized in  Tab \ref{finalresult} and compared to simple HEM-based fusion. Fig \ref{Residualfusion} visualizes the absolute residual maps of TanDEM-X, Cartosat-1, HEM and ANN-based DEM fusion results in comparison to reference data for some exemplary study areas.  

\begin{sidewaystable}
	\centering
	\begin{tabular}{lllll|llll}\hline
		\multicolumn{1}{c}{Areas}&\multicolumn{2}{c}{TanDEM-X}&\multicolumn{2}{c|}{Cartosat-1}&\multicolumn{4}{c}{Fused DEM}\\\hline
		
		&&&&&\multicolumn{2}{c}{HEM}&\multicolumn{2}{c}{ANN}\\
        &NMAD&RMSE&NMAD&RMSE&NMAD&RMSE&NMAD&RMSE\\
		Industrial: D2	    &3.43 & 5.11 &2.43  &3.56 & 2.91 & 4.26 &\cellcolor{green!20}2.42  &\cellcolor{green!20}3.46    \\
		Inner city: I2	    &5.92 & 5.81 &\cellcolor{green!20}4.62  &5.21 & 5.09 & 5.13 &4.90  &\cellcolor{green!20}4.84   \\
		High building: H2	&3.75 & 8.41 &3.29 &8.13 & 3.14 & 7.93  &\cellcolor{green!20}3.13  &\cellcolor{green!20}7.75  \\
		Residential: R2	    &2.30 & 2.61 &2.02  &2.45 & 2.13 & 2.44 &\cellcolor{green!20}1.99  &\cellcolor{green!20}2.33   \\
		Forested: F2     	&3.88 & 4.82 &3.23  &4.65 &3.51  & 4.35 &\cellcolor{green!20}3.19  &\cellcolor{green!20}4.18   \\
		Agricultural: A	    &0.57 & 0.84 &0.65  &0.81 &0.98  & 0.76 &\cellcolor{green!20}0.49  &\cellcolor{green!20}0.68   \\
		\hline	
		
	\end{tabular}
	\caption{Results of fusion of TanDEM-X and Cartosat-1 DEM (in meters) by using weight maps generated by different methods. The weight maps generated by $\dfrac{1}{e_{i}^2}$ and strategy B (see Section \ref{subsec:ANNWeightMap}) were finally adopted for training the ANNs. The green shaded values indicate the best values of metrics relevant to the quality of the compared DEMs.}
	\label{finalresult}
\end{sidewaystable}

\section{Discussion}\label{sec:discussion} 
   
As the obtained NMAD and RMSE results show, standard HEMs generally cannot be reliably used to produce a fused DEM whose height precision exceeds the Cartosat-1 DEM precision. This confirms the assumption that standard HEMs do not reflect all possible error sources in the original DEM data. As an example, the HEM delivered with the TanDEM-X raw DEM just contains error values derived from interferometric coherence and baseline configuration, while deterministic error sources such as layover are not considered. Nevertheless, standard HEMs can be used as a fall-back solution should ground truth for ANN-based weight map prediction be unavailable. 

In contrast, the results obtained for the ANN-supported fusion shows an improvement of the fused DEM product with respect to both input datasets, indicating that the designed ANNs can properly model the existing error patterns related to spatial features that describe the landscaping and the roughness of the land surface under investigation. While the ANN-supported DEM fusion significantly improves the height precision of the TanDEM-X DEM, it also enhances the quality of the Cartosat-1 DEM: As an additional analysis reveals, more than 51\% of all fused DEM pixels are more accurate than their Cartosat-1 counterparts.

\begin{figure}[ht!]
	\begin{center}
		\includegraphics[width=0.8\textwidth]{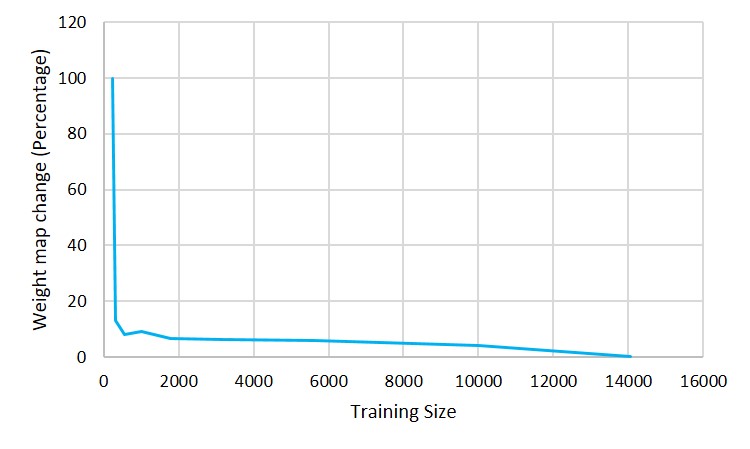}
		\caption{The change of weight map (percentage) according to change in size of training data}
		\label{train-NN}
	\end{center}
\end{figure}

As can be seen in Tab. \ref{studyareades},  the size of training subsets is usually at least as large as the size of the corresponding target subsets, which could lead to the misconception that the number of training and test samples needs to be similar necessarily. In order to show that this is not the case, we conducted an exemplary experiment evaluating the impact of the  training dataset size for the inner city subset. Figure \ref{train-NN} illustrates the influence  of the training sample number on the ANN-predicted weights. Since the actual DEM fusion results depend only on those weights, it can be seen that as few as about 2000 training samples already provide a stable weight prediction and thus stable fusion results. This is due to using a shallow ANN architecture with only a couple of neurons, which produces stable predictions already with a limited amount of training data. This is also supported by  the preprocessing phase described in Section \ref{subsec:Preprocess}. The experiment confirms that the sizes of training and target areas need not be similar for our method to work.  Nevertheless, for the preprocessing and feature extraction , as well as for providing validation data during the training, patch-shaped subsets should be selected instead of selecting only few independent DEM pixels.

Last but not least, the absolute vertical accuracy of the fused DEM is also better than the absolute accuracy of the original Cartosat-1 DEM, which is achieved through the alignment to the more accurately localized TanDEM-X DEM. Thus, eventually, the proposed DEM fusion is able to provide a final DEM product that provides a higher quality than the individual input DEMs in both absolute and relative measures.

 \section{Evaluation of the ANN-based fusion algorithm using other DEM data}\label{Extensionfusion}
 
\begin{figure}[htb]
 	\begin{minipage}[b]{0.9\linewidth}
 		\centering
 		\centerline{\epsfig{figure=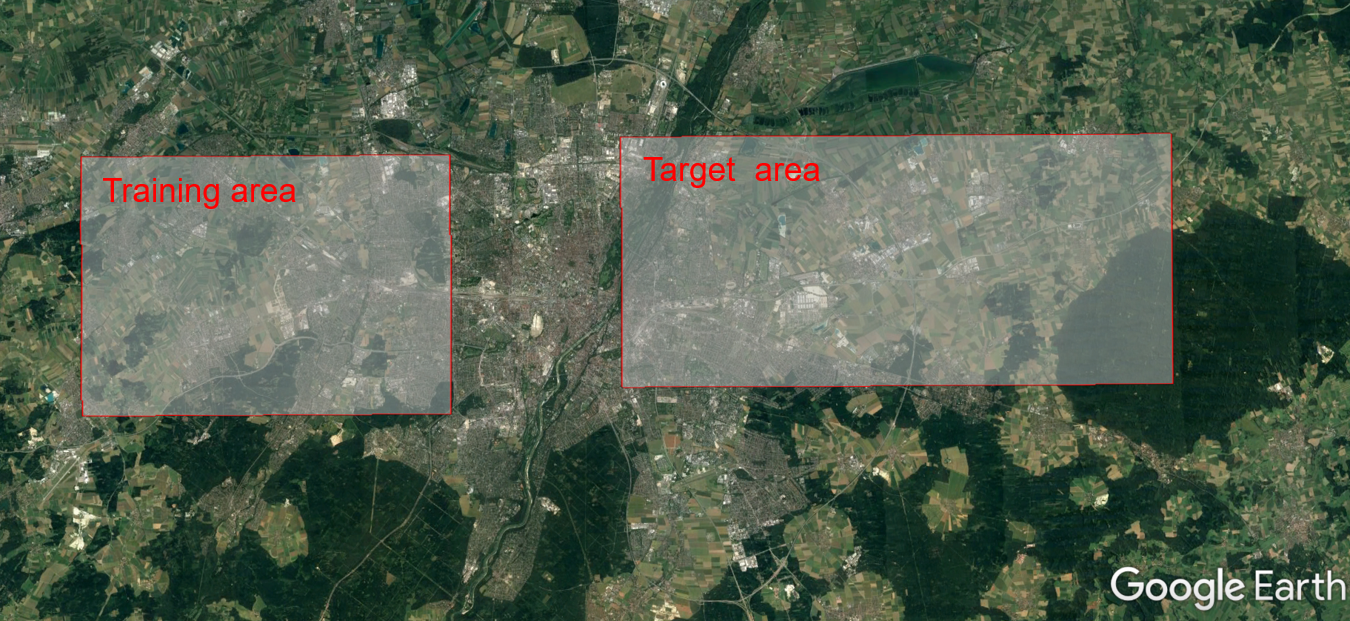,width=1\columnwidth}}
 	\end{minipage}	
 	\caption{Location of training and target areas for ASTER and SRTM DEM fusion study}
 	\label{astersrtm_studyarea}
 \end{figure}

 While the main focus of this study is to fuse the TanDEM-X and Cartosat-1 DEMs according to the reasons mentioned in Section \ref{sec:Introduction}, we also seek to validate the proposed framework for the fusion of other global DEM data. Thus, we carried out similar experiments for ASTER GDEM and SRTM-C DEMs over the Munich area. Fig \ref{astersrtm_studyarea} displays both the training and target areas including the same different urban and sub-urban land types. SRTM data with pixel spacing of 1 arc second (about 30 m) from the C-band data-takes by Shuttle Radar Topography Mission (operated in February 2000) are globally available by USGS portal as InSAR DEMs \cite{USGSSRTM}. In this experiment the void filled SRTM DEM was used. Moreover, version 002 of the ASTER GDEMs was produced from thermal ASTER (Advanced Spaceborne Thermal Emission and Reflection Radiometer) sensor (initially launched in 1999) by stereoscopic 3D reconstruction \cite{USGSASTER}. In addition,  the Cartosat-1 data are used as a ground truth both for purposes of height residual estimation for training of the ANN and assessing the final fusion chain.
 
As we illustrated in Section \ref{sec:fusionresult}, it is not so necessary to take care of strategies A or B for training or different ways of weighting, but for making the process of DEM fusion easier, following strategy B can cope with the task of DEM fusion. As a result, ANN-based fusion according to strategy B was employed for the new DEM data fusion. The fusion results are expressed in Tab \ref{ASTERSRTMresult}. Because no weight maps were provided with the elevation data, the simple averaging approach was used instead of WA  for comparison of the results with our proposed method. The results confirm the efficiency of the proposed algorithm for new cases of DEM fusion as well. It can be clearly inferred from the results that simple averaging just improves the accuracy of the ASTER DEM, and RMSE value of fused DEM does not exceed the RMSE of SRTM DEM, while by using the ANN and following the invented framework in this study, the final RMSE of fused DEM becomes better than initial study DEMs\textemdash meaning the more reliable DEM can be obtained by NNs. In terms of NMAD metric, the quality of ANN-based DEM fusion is significantly higher than simple averaging. 
 
\begin{sidewaystable}
	\centering
	\begin{tabular}{lllll|llll}\hline
		\multicolumn{1}{c}{}&\multicolumn{2}{c}{SRTM-C}&\multicolumn{2}{c|}{ASTER GDEM}&\multicolumn{4}{c}{Fused DEM}\\\hline
		
		&&&&&\multicolumn{2}{c}{Averaging}&\multicolumn{2}{c}{ANN}\\
		&NMAD&RMSE&NMAD&RMSE&NMAD&RMSE&NMAD&RMSE\\
		Target Area   &2.25	&4.01	&4.92	&6.51	&3.14	&4.51	&2.34	&3.89  \\
		\hline
		
	\end{tabular}
	\caption{ASTER GDEM and SRTM-C data fusion results over Munich area using the proposed NN-based approach}
	\label{ASTERSRTMresult}
\end{sidewaystable}

\section{Conclusion}\label{sec:conclusion}

This study focused on the fusion of TanDEM-X and Cartosat-1 DEM over urban areas and their surroundings. The main objective of the investigation was to ultimately obtain the final fused DEM with higher height precision and absolute accuracy. For this task, a typical data fusion technique, weighted averaging, is employed to pixel-wise fuse the elevation data. To achieve optimal fusion results, an innovative framework was developed to predict weight maps using a fully connected artificial neural network. The results demonstrated that the proposed method can efficiently improve the height precision of both Cartosat-1 and TanDEM-X DEMs up to 50 \% in urban areas and 22 \% in non-urban areas as well as the absolute accuracy could be increased through the data alignment. While the height precision improvement by the HEM-based method does not exceed 20\% and 10\% in urban and non-urban areas respectively if HEMs would be available for DEM fusion. Finally, for validating the application of the NN-based approach in other cases of InSAR and optical DEM fusions, different data such as the ASTER GDEM and the SRTM-C elevation data were fused in this way. The results again proved the efficiency of the proposed algorithm for optical and InSAR DEM fusion applications.                  

\section*{Acknowledgment}\label{ACKNOWLEDGEMENTS}
The authors would like to thank Prof.  Reinartz and Dr. \'{d}Angelo of DLR for providing the Cartosat-1 DEM; Dr. Fritz of DLR for providing the TanDEM-X raw DEM; and the Bavarian Surveying Administration for providing the LiDAR reference data. This work is jointly supported by the European Research Council (ERC) under the European Union Horizon 2020 research and innovation program (grant agreement No [ERC-2016-StG-714087], Acronym: \textit{So2Sat}), and Helmholtz Association under the framework of the Young Investigators Group ''SiPEO'' (VH-NG-1018, www.sipeo.bgu.tum.de).

\bibliographystyle{model1-num-names}
\bibliography{sample.bib}

\begin{thebibliography}{45}
\expandafter\ifx\csname natexlab\endcsname\relax\def\natexlab#1{#1}\fi
\providecommand{\bibinfo}[2]{#2}
\ifx\xfnm\relax \def\xfnm[#1]{\unskip,\space#1}\fi
\bibitem[{Rabus et~al.(2003)Rabus, Eineder, Roth, and Bamler}]{Rabus2003}
\bibinfo{author}{B.~Rabus}, \bibinfo{author}{M.~Eineder},
  \bibinfo{author}{A.~Roth}, \bibinfo{author}{R.~Bamler},
\newblock \bibinfo{title}{The shuttle radar topography mission—a new class of
  digital elevation models acquired by spaceborne radar},
\newblock \bibinfo{journal}{ISPRS Journal of Photogrammetry and Remote Sensing}
  \bibinfo{volume}{57} (\bibinfo{year}{2003}) \bibinfo{pages}{241 -- 262}.
\bibitem[{Rodriguez et~al.(2006)Rodriguez, Morris, and Belz}]{Rodriguez2006}
\bibinfo{author}{E.~Rodriguez}, \bibinfo{author}{C.~S. Morris},
  \bibinfo{author}{J.~E. Belz},
\newblock \bibinfo{title}{A global assessment of the {SRTM} performance},
\newblock \bibinfo{journal}{Photogrammetric Engineering \& Remote Sensing}
  \bibinfo{volume}{72} (\bibinfo{year}{2006}) \bibinfo{pages}{249--260}.
\bibitem[{Tachikawa et~al.(2011)Tachikawa, Kaku, Iwasaki, Gesch, Oimoen, Zhang,
  Danielson, Krieger, Curtis, Haase et~al.}]{Tachikawa2011}
\bibinfo{author}{T.~Tachikawa}, \bibinfo{author}{M.~Kaku},
  \bibinfo{author}{A.~Iwasaki}, \bibinfo{author}{D.~B. Gesch},
  \bibinfo{author}{M.~J. Oimoen}, \bibinfo{author}{Z.~Zhang},
  \bibinfo{author}{J.~J. Danielson}, \bibinfo{author}{T.~Krieger},
  \bibinfo{author}{B.~Curtis}, \bibinfo{author}{J.~Haase}, et~al.,
  \bibinfo{title}{{ASTER} global digital elevation model version 2-summary of
  validation results}, \bibinfo{type}{Technical Report}, NASA,
  \bibinfo{year}{2011}.
\bibitem[{Takaku et~al.(2014)Takaku, Tadono, and
  Tsutsui}]{takaku2014generation}
\bibinfo{author}{J.~Takaku}, \bibinfo{author}{T.~Tadono},
  \bibinfo{author}{K.~Tsutsui},
\newblock \bibinfo{title}{Generation of high resolution global {DSM} from {ALOS
  PRISM}},
\newblock \bibinfo{journal}{The International Archives of Photogrammetry,
  Remote Sensing and Spatial Information Sciences} \bibinfo{volume}{40}
  (\bibinfo{year}{2014}) \bibinfo{pages}{243}.
\bibitem[{Tadono et~al.(2014)Tadono, Ishida, Oda, Naito, Minakawa, and
  Iwamoto}]{tadono2014precise}
\bibinfo{author}{T.~Tadono}, \bibinfo{author}{H.~Ishida},
  \bibinfo{author}{F.~Oda}, \bibinfo{author}{S.~Naito},
  \bibinfo{author}{K.~Minakawa}, \bibinfo{author}{H.~Iwamoto},
\newblock \bibinfo{title}{Precise global {DEM} generation by {ALOS PRISM}},
\newblock \bibinfo{journal}{ISPRS Annals of the Photogrammetry, Remote Sensing
  and Spatial Information Sciences} \bibinfo{volume}{2} (\bibinfo{year}{2014})
  \bibinfo{pages}{71}.
\bibitem[{Takaku et~al.(2016)Takaku, Tadono, Tsutsui, and
  Ichikawa}]{takaku2016validation}
\bibinfo{author}{J.~Takaku}, \bibinfo{author}{T.~Tadono},
  \bibinfo{author}{K.~Tsutsui}, \bibinfo{author}{M.~Ichikawa},
\newblock \bibinfo{title}{Validation of {" AW3D"} global {DSM} generated from
  {ALOS PRISM}},
\newblock \bibinfo{journal}{ISPRS Annals of the Photogrammetry, Remote Sensing
  and Spatial Information Sciences} \bibinfo{volume}{3} (\bibinfo{year}{2016})
  \bibinfo{pages}{25}.
\bibitem[{Heady et~al.(2009)Heady, Kroenung, and Rodarmel}]{heady2009high}
\bibinfo{author}{B.~Heady}, \bibinfo{author}{G.~Kroenung},
  \bibinfo{author}{C.~Rodarmel},
\newblock \bibinfo{title}{High resolution elevation data ({HRE}) specification
  overview},
\newblock in: \bibinfo{booktitle}{ASPRS/MAPPS 2009 Conference, San Antonio,
  Texas}.
\bibitem[{Krieger et~al.(2007)Krieger, Moreira, Fiedler, Hajnsek, Werner,
  Younis, and Zink}]{Krieger2007}
\bibinfo{author}{G.~Krieger}, \bibinfo{author}{A.~Moreira},
  \bibinfo{author}{H.~Fiedler}, \bibinfo{author}{I.~Hajnsek},
  \bibinfo{author}{M.~Werner}, \bibinfo{author}{M.~Younis},
  \bibinfo{author}{M.~Zink},
\newblock \bibinfo{title}{{TanDEM-X:} a satellite formation for high-resolution
  {SAR} interferometry},
\newblock \bibinfo{journal}{IEEE Transactions on Geoscience and Remote Sensing}
  \bibinfo{volume}{45} (\bibinfo{year}{2007}) \bibinfo{pages}{3317--3341}.
\bibitem[{Fritz et~al.(2011)Fritz, Rossi, Yague-Martinez, Rodriguez-Gonzalez,
  Lachaise, and Breit}]{6049701}
\bibinfo{author}{T.~Fritz}, \bibinfo{author}{C.~Rossi},
  \bibinfo{author}{N.~Yague-Martinez}, \bibinfo{author}{F.~Rodriguez-Gonzalez},
  \bibinfo{author}{M.~Lachaise}, \bibinfo{author}{H.~Breit},
\newblock \bibinfo{title}{Interferometric processing of {TanDEM-X} data},
\newblock in: \bibinfo{booktitle}{2011 IEEE International Geoscience and Remote
  Sensing Symposium}, pp. \bibinfo{pages}{2428--2431}.
\bibitem[{Gruber et~al.(2012)Gruber, Wessel, Huber, and Roth}]{Gruber2012}
\bibinfo{author}{A.~Gruber}, \bibinfo{author}{B.~Wessel},
  \bibinfo{author}{M.~Huber}, \bibinfo{author}{A.~Roth},
\newblock \bibinfo{title}{Operational {TanDEM-X DEM} calibration and first
  validation results},
\newblock \bibinfo{journal}{ISPRS Journal of Photogrammetry and Remote Sensing}
  \bibinfo{volume}{73} (\bibinfo{year}{2012}) \bibinfo{pages}{39 -- 49}.
  \bibinfo{note}{Innovative Applications of SAR Interferometry from modern
  Satellite Sensors}.
\bibitem[{Rossi et~al.(2012)Rossi, Gonzalez, Fritz, Yague-Martinez, and
  Eineder}]{Rossi2012}
\bibinfo{author}{C.~Rossi}, \bibinfo{author}{F.~R. Gonzalez},
  \bibinfo{author}{T.~Fritz}, \bibinfo{author}{N.~Yague-Martinez},
  \bibinfo{author}{M.~Eineder},
\newblock \bibinfo{title}{{TanDEM-X} calibrated raw {DEM} generation},
\newblock \bibinfo{journal}{ISPRS Journal of Photogrammetry and Remote Sensing}
  \bibinfo{volume}{73} (\bibinfo{year}{2012}) \bibinfo{pages}{12 -- 20}.
\bibitem[{Gruber et~al.(2016)Gruber, Wessel, Martone, and Roth}]{Gruber2016}
\bibinfo{author}{A.~Gruber}, \bibinfo{author}{B.~Wessel},
  \bibinfo{author}{M.~Martone}, \bibinfo{author}{A.~Roth},
\newblock \bibinfo{title}{The {TanDEM-X DEM} mosaicking: Fusion of multiple
  acquisitions using {InSAR} quality parameters},
\newblock \bibinfo{journal}{IEEE Journal of Selected Topics in Applied Earth
  Observations and Remote Sensing} \bibinfo{volume}{9} (\bibinfo{year}{2016})
  \bibinfo{pages}{1047--1057}.
\bibitem[{Zink et~al.(2014)Zink, Bachmann, Brautigam, Fritz, Hajnsek, Moreira,
  Wessel, and Krieger}]{Zink2014}
\bibinfo{author}{M.~Zink}, \bibinfo{author}{M.~Bachmann},
  \bibinfo{author}{B.~Brautigam}, \bibinfo{author}{T.~Fritz},
  \bibinfo{author}{I.~Hajnsek}, \bibinfo{author}{A.~Moreira},
  \bibinfo{author}{B.~Wessel}, \bibinfo{author}{G.~Krieger},
\newblock \bibinfo{title}{{TanDEM-X:} the new global {DEM} takes shape},
\newblock \bibinfo{journal}{IEEE Geoscience and Remote Sensing Magazine}
  \bibinfo{volume}{2} (\bibinfo{year}{2014}) \bibinfo{pages}{8--23}.
\bibitem[{Rossi and Gernhardt(2013)}]{Rossi2013}
\bibinfo{author}{C.~Rossi}, \bibinfo{author}{S.~Gernhardt},
\newblock \bibinfo{title}{Urban {DEM} generation, analysis and enhancements
  using {TanDEM-X}},
\newblock \bibinfo{journal}{ISPRS Journal of Photogrammetry and Remote Sensing}
  \bibinfo{volume}{85} (\bibinfo{year}{2013}) \bibinfo{pages}{120 -- 131}.
\bibitem[{Rossi et~al.(2011)Rossi, Fritz, Breit, and Eineder}]{5764720}
\bibinfo{author}{C.~Rossi}, \bibinfo{author}{T.~Fritz},
  \bibinfo{author}{H.~Breit}, \bibinfo{author}{M.~Eineder},
\newblock \bibinfo{title}{First urban {TanDEM-X} raw {DEMs} analysis},
\newblock in: \bibinfo{booktitle}{2011 Joint Urban Remote Sensing Event}, pp.
  \bibinfo{pages}{65--68}.
\bibitem[{Srivastava et~al.(2007)Srivastava, Srinivasan, Gupta, Singh, Nain,
  Prakash, Kartikeyan, and Krishna}]{Srivastava2007}
\bibinfo{author}{P.~K. Srivastava}, \bibinfo{author}{T.~Srinivasan},
  \bibinfo{author}{A.~Gupta}, \bibinfo{author}{S.~Singh},
  \bibinfo{author}{J.~S. Nain}, \bibinfo{author}{S.~Prakash},
  \bibinfo{author}{B.~Kartikeyan}, \bibinfo{author}{B.~G. Krishna},
\newblock \bibinfo{title}{Recent advances in {CARTOSAT-1} data processing},
\newblock \bibinfo{journal}{ISPRS - International Archives of the
  Photogrammetry, Remote Sensing and Spatial Information Sciences}
  \bibinfo{volume}{XXXVI-1/W51} (\bibinfo{year}{2007}) \bibinfo{pages}{on
  CD--ROM}.
\bibitem[{Ahmed et~al.(2007)Ahmed, Mahtab, Agrawal, Jayaprasad, Pathan, Ajai,
  Singh, and Singh}]{Ahmed2007}
\bibinfo{author}{N.~Ahmed}, \bibinfo{author}{A.~Mahtab},
  \bibinfo{author}{R.~Agrawal}, \bibinfo{author}{P.~Jayaprasad},
  \bibinfo{author}{S.~K. Pathan}, \bibinfo{author}{Ajai},
  \bibinfo{author}{D.~K. Singh}, \bibinfo{author}{A.~K. Singh},
\newblock \bibinfo{title}{Extraction and validation of {Cartosat-1 DEM}},
\newblock \bibinfo{journal}{Journal of the Indian Society of Remote Sensing}
  \bibinfo{volume}{35} (\bibinfo{year}{2007}) \bibinfo{pages}{121}.
\bibitem[{Uttenthaler et~al.(2013)Uttenthaler, Barner, Hass, Makiola,
  d’Angelo, Reinartz, Carl, and Steiner}]{Uttenthaler2013}
\bibinfo{author}{A.~Uttenthaler}, \bibinfo{author}{F.~Barner},
  \bibinfo{author}{T.~Hass}, \bibinfo{author}{J.~Makiola},
  \bibinfo{author}{P.~d’Angelo}, \bibinfo{author}{P.~Reinartz},
  \bibinfo{author}{S.~Carl}, \bibinfo{author}{K.~Steiner},
\newblock \bibinfo{title}{{EURO-MAPS 3D-} {A} transnational, high-resolution
  digital surface model for {Europe}},
\newblock in: \bibinfo{booktitle}{ESA Living Planet Symposium}, volume
  \bibinfo{volume}{722}, p. \bibinfo{pages}{271}.
\bibitem[{Lehner et~al.(2007)Lehner, M{\"u}ller, Reinartz, and
  Schroeder}]{Lehner2007}
\bibinfo{author}{M.~Lehner}, \bibinfo{author}{R.~M{\"u}ller},
  \bibinfo{author}{P.~Reinartz}, \bibinfo{author}{M.~Schroeder},
\newblock \bibinfo{title}{Stereo evaluation of {Cartosat-1} data for {French}
  and {Catalonian} test sites},
\newblock \bibinfo{journal}{ISPRS - International Archives of the
  Photogrammetry, Remote Sensing and Spatial Information Sciences}
  \bibinfo{volume}{XXXVI-1/W51} (\bibinfo{year}{2007}) \bibinfo{pages}{on
  CD--ROM}.
\bibitem[{Teo(2011)}]{Teo2011}
\bibinfo{author}{T.-A. Teo},
\newblock \bibinfo{title}{Bias compensation in a rigorous sensor model and
  rational function model for high-resolution satellite images},
\newblock \bibinfo{journal}{Photogrammetric Engineering \& Remote Sensing}
  \bibinfo{volume}{77} (\bibinfo{year}{2011}) \bibinfo{pages}{1211--1220}.
\bibitem[{Kim and Jeong(2011)}]{Kim2011}
\bibinfo{author}{T.~Kim}, \bibinfo{author}{J.~Jeong},
\newblock \bibinfo{title}{{DEM} matching for bias compensation of rigorous
  pushbroom sensor models},
\newblock \bibinfo{journal}{ISPRS Journal of Photogrammetry and Remote Sensing}
  \bibinfo{volume}{66} (\bibinfo{year}{2011}) \bibinfo{pages}{692 -- 699}.
\bibitem[{Schmitt and Zhu(2016)}]{7740215}
\bibinfo{author}{M.~Schmitt}, \bibinfo{author}{X.~X. Zhu},
\newblock \bibinfo{title}{Data fusion and remote sensing: An ever-growing
  relationship},
\newblock \bibinfo{journal}{IEEE Geoscience and Remote Sensing Magazine}
  \bibinfo{volume}{4} (\bibinfo{year}{2016}) \bibinfo{pages}{6--23}.
\bibitem[{Reinartz et~al.(2005)Reinartz, M{\"u}ller, Hoja, Lehner, and
  Schroeder}]{Reinartz2005}
\bibinfo{author}{P.~Reinartz}, \bibinfo{author}{R.~M{\"u}ller},
  \bibinfo{author}{D.~Hoja}, \bibinfo{author}{M.~Lehner},
  \bibinfo{author}{M.~Schroeder},
\newblock \bibinfo{title}{Comparison and fusion of {DEM} derived from {SPOT-5
  HRS} and {SRTM} data and estimation of forest heights},
\newblock in: \bibinfo{editor}{Earsel} (Ed.), \bibinfo{booktitle}{Earsel
  Symposium, Porto, Portugal, 6.-11. June 2005}.
\bibitem[{Roth et~al.(2002)Roth, Knopfle, Strunz, Lehner, and
  Reinartz}]{Roth2002}
\bibinfo{author}{A.~Roth}, \bibinfo{author}{W.~Knopfle},
  \bibinfo{author}{G.~Strunz}, \bibinfo{author}{M.~Lehner},
  \bibinfo{author}{P.~Reinartz},
\newblock \bibinfo{title}{Towards a global elevation product: combination of
  multi-source digital elevation models},
\newblock \bibinfo{journal}{International Archives of Photogrammetry Remote
  Sensing and Spatial Information Sciences} \bibinfo{volume}{34}
  (\bibinfo{year}{2002}) \bibinfo{pages}{675--679}.
\bibitem[{Papasaika et~al.(2011)Papasaika, Kokiopoulou, Baltsavias, Schindler,
  and Kressner}]{Papasaika:2011:FDE:2050390.2050409}
\bibinfo{author}{H.~Papasaika}, \bibinfo{author}{E.~Kokiopoulou},
  \bibinfo{author}{E.~Baltsavias}, \bibinfo{author}{K.~Schindler},
  \bibinfo{author}{D.~Kressner},
\newblock \bibinfo{title}{Fusion of digital elevation models using sparse
  representations},
\newblock in: \bibinfo{booktitle}{Proceedings of the 2011 ISPRS Conference on
  Photogrammetric Image Analysis}, PIA'11,
  \bibinfo{publisher}{Springer-Verlag}, \bibinfo{address}{Berlin, Heidelberg},
  \bibinfo{year}{2011}, pp. \bibinfo{pages}{171--184}.
\bibitem[{Pock et~al.(2011)Pock, Zebedin, and Bischof}]{Pock2011}
\bibinfo{author}{T.~Pock}, \bibinfo{author}{L.~Zebedin},
  \bibinfo{author}{H.~Bischof}, \bibinfo{title}{{TGV}-Fusion},
  \bibinfo{publisher}{Springer Berlin Heidelberg}, \bibinfo{address}{Berlin,
  Heidelberg}, pp. \bibinfo{pages}{245--258}.
\bibitem[{Kuschk et~al.(2017)Kuschk, d’Angelo, Gaudrie, Reinartz, and
  Cremers}]{7752839}
\bibinfo{author}{G.~Kuschk}, \bibinfo{author}{P.~d’Angelo},
  \bibinfo{author}{D.~Gaudrie}, \bibinfo{author}{P.~Reinartz},
  \bibinfo{author}{D.~Cremers},
\newblock \bibinfo{title}{Spatially regularized fusion of multiresolution
  digital surface models},
\newblock \bibinfo{journal}{IEEE Transactions on Geoscience and Remote Sensing}
  \bibinfo{volume}{55} (\bibinfo{year}{2017}) \bibinfo{pages}{1477--1488}.
\bibitem[{Fuss et~al.(2016)Fuss, Berg, and Lindsay}]{Fuss2016}
\bibinfo{author}{C.~E. Fuss}, \bibinfo{author}{A.~A. Berg},
  \bibinfo{author}{J.~B. Lindsay},
\newblock \bibinfo{title}{Dem fusion using a modified k-means clustering
  algorithm},
\newblock \bibinfo{journal}{International Journal of Digital Earth}
  \bibinfo{volume}{9} (\bibinfo{year}{2016}) \bibinfo{pages}{1242--1255}.
\bibitem[{Rossi et~al.(2013)Rossi, Eineder, Fritz, d'Angelo, and
  Reinartz}]{dlr87859}
\bibinfo{author}{C.~Rossi}, \bibinfo{author}{M.~Eineder},
  \bibinfo{author}{T.~Fritz}, \bibinfo{author}{P.~d'Angelo},
  \bibinfo{author}{P.~Reinartz},
\newblock \bibinfo{title}{Quality assessment of {TanDEM-X} raw {DEMs} oriented
  to a fusion with {CartoSAT-1 DEMs}},
\newblock in: \bibinfo{booktitle}{33rd EARSeL Symposium}, pp.
  \bibinfo{pages}{1--9}.
\bibitem[{Bagheri et~al.(2017)Bagheri, Schmitt, and Zhu}]{Bagheri2017}
\bibinfo{author}{H.~Bagheri}, \bibinfo{author}{M.~Schmitt},
  \bibinfo{author}{X.~X. Zhu},
\newblock \bibinfo{title}{Uncertainty assessment and weight map generation for
  efficient fusion of {TanDEM-X} and {Cartosat-1 DEMs}},
\newblock \bibinfo{journal}{ISPRS - International Archives of the
  Photogrammetry, Remote Sensing and Spatial Information Sciences}
  \bibinfo{volume}{XLII-1/W1} (\bibinfo{year}{2017}) \bibinfo{pages}{433--439}.
\bibitem[{Deo et~al.(2015)Deo, Rossi, Eineder, Fritz, and Rao}]{Deo2015}
\bibinfo{author}{R.~Deo}, \bibinfo{author}{C.~Rossi},
  \bibinfo{author}{M.~Eineder}, \bibinfo{author}{T.~Fritz},
  \bibinfo{author}{Y.~S. Rao},
\newblock \bibinfo{title}{Framework for fusion of ascending and descending pass
  {TanDEM-X} raw dems},
\newblock \bibinfo{journal}{IEEE Journal of Selected Topics in Applied Earth
  Observations and Remote Sensing} \bibinfo{volume}{8} (\bibinfo{year}{2015})
  \bibinfo{pages}{3347--3355}.
\bibitem[{Bagheri et~al.(2017)Bagheri, Schmitt, and Zhu}]{bagheri2017fusion}
\bibinfo{author}{H.~Bagheri}, \bibinfo{author}{M.~Schmitt},
  \bibinfo{author}{X.~X. Zhu},
\newblock \bibinfo{title}{Fusion of {TanDEM-X} and {Cartosat-1} {DEMs} using
  {TV-}norm regularization and {ANN-}predicted weights},
\newblock in: \bibinfo{booktitle}{Proceedings of IEEE Geosci. and Remote Sens.
  Symposium}.
\bibitem[{d'Angelo et~al.(2008)d'Angelo, Lehner, Krauss, Hoja, and
  Reinartz}]{dlr55978}
\bibinfo{author}{P.~d'Angelo}, \bibinfo{author}{M.~Lehner},
  \bibinfo{author}{T.~Krauss}, \bibinfo{author}{D.~Hoja},
  \bibinfo{author}{P.~Reinartz},
\newblock \bibinfo{title}{Towards automated {DEM} generation from high
  resolution stereo satellite images},
\newblock \bibinfo{journal}{ISPRS - International Archives of the
  Photogrammetry, Remote Sensing and Spatial Information Sciences}
  \bibinfo{volume}{XXXVII} (\bibinfo{year}{2008}) \bibinfo{pages}{1137--1342}.
\bibitem[{{The Federal Agency for Cartography and Geodesy of Germany
  (BKG)}(2016)}]{Cartography}
\bibinfo{author}{{The Federal Agency for Cartography and Geodesy of Germany
  (BKG)}}, \bibinfo{title}{Digital orthophotos},
  \bibinfo{howpublished}{\url{https://www.bkg.bund.de/SharedDocs/Downloads/BKG/DE/Downloads-DE-Flyer/AdV-DOP-DE}},
  \bibinfo{year}{2016}. \bibinfo{note}{(Accessed 09.17)}.
\bibitem[{Ravanbakhsh and Fraser(2013)}]{Ravanbakhsh2013}
\bibinfo{author}{M.~Ravanbakhsh}, \bibinfo{author}{C.~S. Fraser},
\newblock \bibinfo{title}{A comparative study of {DEM} registration
  approaches},
\newblock \bibinfo{journal}{Journal of Spatial Science} \bibinfo{volume}{58}
  (\bibinfo{year}{2013}) \bibinfo{pages}{79--89}.
\bibitem[{Martone et~al.(2012)Martone, Bräutigam, Rizzoli, Gonzalez, Bachmann,
  and Krieger}]{Martone2012}
\bibinfo{author}{M.~Martone}, \bibinfo{author}{B.~Bräutigam},
  \bibinfo{author}{P.~Rizzoli}, \bibinfo{author}{C.~Gonzalez},
  \bibinfo{author}{M.~Bachmann}, \bibinfo{author}{G.~Krieger},
\newblock \bibinfo{title}{Coherence evaluation of {TanDEM-X} interferometric
  data},
\newblock \bibinfo{journal}{ISPRS Journal of Photogrammetry and Remote Sensing}
  \bibinfo{volume}{73} (\bibinfo{year}{2012}) \bibinfo{pages}{21 -- 29}.
\bibitem[{Just and Bamler(1994)}]{Just1994}
\bibinfo{author}{D.~Just}, \bibinfo{author}{R.~Bamler},
\newblock \bibinfo{title}{Phase statistics of interferograms with applications
  to synthetic aperture radar},
\newblock \bibinfo{journal}{Applied optics} \bibinfo{volume}{33}
  (\bibinfo{year}{1994}) \bibinfo{pages}{4361—4368}.
\bibitem[{Toutin(2002)}]{Toutin2002}
\bibinfo{author}{T.~Toutin},
\newblock \bibinfo{title}{Impact of terrain slope and aspect on radargrammetric
  {DEM} accuracy},
\newblock \bibinfo{journal}{ISPRS Journal of Photogrammetry and Remote Sensing}
  \bibinfo{volume}{57} (\bibinfo{year}{2002}) \bibinfo{pages}{228 -- 240}.
\bibitem[{Papasaika et~al.(2009)Papasaika, Poli, and Baltsavias}]{4782702}
\bibinfo{author}{H.~Papasaika}, \bibinfo{author}{D.~Poli},
  \bibinfo{author}{E.~Baltsavias},
\newblock \bibinfo{title}{Fusion of digital elevation models from various data
  sources},
\newblock in: \bibinfo{booktitle}{2009 International Conference on Advanced
  Geographic Information Systems Web Services}, pp. \bibinfo{pages}{117--122}.
\bibitem[{Reinartz et~al.(2010)Reinartz, d'Angelo, Krau{\ss}, Poli, Jacobsen,
  and Buyuksalih}]{Reinartz2010}
\bibinfo{author}{P.~Reinartz}, \bibinfo{author}{P.~d'Angelo},
  \bibinfo{author}{T.~Krau{\ss}}, \bibinfo{author}{D.~Poli},
  \bibinfo{author}{K.~Jacobsen}, \bibinfo{author}{G.~Buyuksalih},
\newblock \bibinfo{title}{Benchmarking and quality analysis of {DEM} generated
  from high and very high resolution optical stereo satellite data},
\newblock \bibinfo{journal}{ISPRS - International Archives of the
  Photogrammetry, Remote Sensing and Spatial Information Sciences}
  \bibinfo{volume}{XXXVIII} (\bibinfo{year}{2010}).
\bibitem[{Olaya(2009)}]{Olaya2009c}
\bibinfo{author}{V.~Olaya},
\newblock \bibinfo{title}{{Chapter 6: Basic Land-Surface Parameters}},
\newblock in: \bibinfo{editor}{T.~H. Science}, \bibinfo{editor}{H.~I. R. B.
  T.~D. in~Soil} (Eds.), \bibinfo{booktitle}{GeomorphometryConcepts, Software,
  Applications}, volume~\bibinfo{volume}{33}, \bibinfo{publisher}{Elsevier},
  \bibinfo{year}{2009}, pp. \bibinfo{pages}{141--169}.
\bibitem[{H\"ohle and H\"ohle(2009)}]{Hoehle2009}
\bibinfo{author}{J.~H\"ohle}, \bibinfo{author}{M.~H\"ohle},
\newblock \bibinfo{title}{Accuracy assessment of digital elevation models by
  means of robust statistical methods},
\newblock \bibinfo{journal}{ISPRS Journal of Photogrammetry and Remote Sensing}
  \bibinfo{volume}{64} (\bibinfo{year}{2009}) \bibinfo{pages}{398 -- 406}.
\bibitem[{Birgé and Rozenholc(2006)}]{Birge2006}
\bibinfo{author}{L.~Birgé}, \bibinfo{author}{Y.~Rozenholc},
\newblock \bibinfo{title}{How many bins should be put in a regular histogram},
\newblock \bibinfo{journal}{ESAIM: Probability and Statistics}
  \bibinfo{volume}{10} (\bibinfo{year}{2006}) \bibinfo{pages}{24–45}.
\bibitem[{USGS(2000)}]{USGSSRTM}
\bibinfo{author}{USGS}, \bibinfo{title}{{S}huttle {R}adar {T}opography
  {M}ission ({SRTM}) void filled},
  \bibinfo{howpublished}{\url{https://lta.cr.usgs.gov/SRTMVF}},
  \bibinfo{year}{2000}. \bibinfo{note}{(accessed 09.17)}.
\bibitem[{USGS(1999)}]{USGSASTER}
\bibinfo{author}{USGS}, \bibinfo{title}{Routine {ASTER} global digital
  elevation model},
  \bibinfo{howpublished}{\url{https://lpdaac.usgs.gov/dataset_discovery/aster/aster_products_table/astgtm}},
  \bibinfo{year}{1999}.

\end{thebibliography}

\end{document}